%% file: dust_multi.tex
\documentclass[a4paper,usenatbib,times,fleqn]{mn2e}
\usepackage{natbib,graphicx,amsmath,amsfonts,amssymb,times,txfonts,pstricks,multicol}

\def\vg{\textbf{v}_{\mathrm{g}}}
\def\vd{\textbf{v}_{\mathrm{d}}}
\def\vdk{\textbf{v}_{\mathrm{d}k}}
\def\rhog{\rho_{\mathrm{g}}}
\def\rhod{\rho_{\mathrm{d}}}
\def\rhodk{\rho_{\mathrm{d}k}}

\def\tk{t_{k}}
\def\ts{t_{\mathrm{s}}}

\def\deltav{\Delta \textbf{v}}
\def\deltavk{\Delta \textbf{v}_{k}}
\def\deltavkx{\Delta v_{kx}}
\def\deltavky{\Delta v_{ky}}
\def\deltavix{\Delta v_{ix}}
\def\deltaviy{\Delta v_{iy}}

\def\deltavl{\Delta \textbf{v}_{l}}
\def\cs{c_{\mathrm{s}}}
\def\vb{\textbf{v}}

\def\rhoz{\rho_{0}}

\def\ge{g_{\mathrm{e}}}

\def\fg{\mathbf{f}_{\mathrm{g}}}
\def\fd{\mathbf{f}_{\mathrm{d}}}
\def\fdk{\mathbf{f}_{\mathrm{d}k}}

\def\epsd{\epsilon}

\def\epsz{\epsilon_{0}}

\def\vk{\textbf{v}_{k}}

\def\sk{\sum_{k}}

\def\sl{\sum_{l}}
\def\kk{K_{k}}
\def\epsk{\epsilon_{k}}
\def\epskz{\epsilon_{k0}}
\def\epsl{\epsilon_{l}}
\def\epslz{\epsilon_{l0}}

\def\tbk{t_{\mathrm{b}k}}

\def\tbl{t_{\mathrm{b}l}}

\def\deltabv{\Delta \mathbf{V}}
\def\deltabvh{\Delta \mathbf{\hat{V}}}
\def\deltabvt{\Delta \mathbf{\tilde{V}}}
\def\deltabvi{\Delta V_{i}}
\def\deltabvti{\Delta \tilde{V}_{i}}
\def\Odn{\mathrm{\Omega}_{n}}
\def\Wdn{\mathrm{W}_{n}}
\def\tse{t_{\mathrm{s,eff}}}
\def\teps{\tilde{\epsilon}}
\def\ds{\mathrm{d}s}
\def\dsp{\mathrm{d}s'}

\def\dst{\displaystyle}

\defcitealias{LP12a}{LP12a}
\defcitealias{LP12b}{LP12b}
\defcitealias{LP14a}{LP14a}
\defcitealias{LP14b}{LP14b}
\defcitealias{NSH86}{NSH86}

\title[]{Dust and gas mixtures with multiple grain species --- a one-fluid approach}

\author[Laibe \& Price]{Guillaume Laibe$^{1}$ and Daniel J. Price$^{2}$ \\
$^{1}$School of Physics and Astronomy, University of St. Andrews, North Haugh, St. Andrews, Fife KY16 9SS, UK \\
$^{2}$Monash Centre for Astrophysics and School of Mathematical Sciences, Monash University, Clayton, Vic 3800, Australia
}
\pagerange{\pageref{firstpage}--\pageref{lastpage}} \pubyear{2014}

\begin{document}
\include{journaux}

\label{firstpage}
\bibliographystyle{mn2e}
\maketitle

\begin{abstract}
We derive the single-fluid evolution equations describing a mixture made of a gas phase and an arbitrary number of dust phases, generalising the approach developed in \citet{LP14a}. A generalisation for continuous dust distributions as well as analytic approximations for strong drag regimes are also provided. This formalism lays the foundation for numerical simulations of dust populations in a wide range of astrophysical systems while avoiding limitations associated with a multiple-fluid treatment.

The usefulness of the formalism is illustrated on a series of analytical problems, namely the \textsc{dustybox}, \textsc{dustyshock} and \textsc{dustywave} problems as well as the radial drift of grains and the streaming instability in protoplanetary discs. We find physical effects specific to the presence of several dust phases and multiple drag timescales, including non-monotonic evolution of the differential velocity between phases and increased efficiency of the linear growth of the streaming instability. Interestingly, it is found that under certain conditions, large grains can migrate outwards in protoplanetary discs. This may explain the presence of small pebbles at several hundreds of astronomical units from their central star.

\end{abstract}


%----------------------------------------------------------------------------------------------------------------
\section{Introduction}
\label{sec:intro}

Small but not insignificant: Dust grains play an essential role for forming stars and planets in the Universe (e.g.  \citealt{ChiangYoudin2010,Testi2014}). Dust reprocesses the energy emitted from surrounding stars and grains grow to build large solid bodies. Dust in molecular clouds originates from the interstellar medium, where grains have a typical distribution in size of the form $n(s)\propto s^{-3.5}$ \citep{Mathis1977}. Evidence of multiple grain size populations has also been detected in molecular clouds (e.g. \citealt{Pagani2010,Andersen2013}) and in protoplanetary discs (e.g. \citealt{Dullemond04,Duchene04,Pinte07,Lommen2009,Banzatti2011,Ubach2012}). Since the coupling efficiency with the surrounding gas varies with the particle size, different grain populations may experience very different dynamics (e.g. \citealt{Shariff2009}).

Dust evolution has been studied in astrophysical systems mostly by modelling the dust phase as a continuous pressureless fluid and treating the interactions with the gas via a drag force (e.g. \citealt{Saffman1962,GaraudLin2004}). However, numerical simulations using this two fluid approach suffer from two severe limitations \citep{LP12a,LP12b}. Firstly, grain collisions are generally not effective enough to provide support against dust accumulation. Hence, if grains concentrate below the gas resolution (as during the planet formation process), they form dead artificial clumps. Secondly, the presence of small grains requires the use of a prohibitively high spatial resolution in order to resolve the tiny spatial de-phasing of the two phases. These difficulties limit progress in simulating complex dust evolution in cold astrophysical systems, in particular the formation of a planet \textit{ab initio}.

In \citet{LP14a}, we have shown that these limitations can be overcome by changing the physical description of the system, describing the gas and the dust particles as the elementary constituents of a single fluid --- the mixture --- whose density is the total density of its two phases and which is advected at the barycentric velocity of the particles. The chemical composition of the system and the relative velocities between the phases are treated as internal properties of the mixture. Using this description, the fundamental difficulties described above disappear, as shown in our numerical simulations based on this approach using Smoothed Particle Hydrodynamics (SPH) \citep{LP14b}. A single resolution length is used in the simulation, meaning that one phase cannot accumulate below the resolution of the other. Moreover, the resolution criterion arising from the spatial dephasing between the two phases is no longer necessary in this description. Finally, no interpolation between the gas and the dust phases is required and implicit timestepping is straightforward to implement.

The main limitation of the \citet{LP14a,LP14b} work is that only a single dust grain population was considered. This is insufficient for modelling systems where grains of different sizes mix. For example, a good knowledge of the dust distribution is required to compute opacities in radiation-hydrodynamics simulations of star formation. In this paper, we generalise our previous work to describe a mixture of $n$ dust species interacting with a gas component. The equations are given in their most general form in Sect.~\ref{sec:onefluid}. In Sect.~\ref{sec:appli}, the physical properties of multiple dust population mixtures are discussed by applying the one-fluid formalism to analytical examples relevant to astrophysics. In doing so, we provide analytic solutions that can be used to benchmark numerical implementations and which shed light on the rich physics of multiple dust-phase mixtures.

%=======================================================================================================
\section{One fluid with multiple dust species}
\label{sec:onefluid}

We address the problem of treating a mixture composed by a continuous gas phase and any number $n$ of distinct dust phases (e.g. made of different grain sizes). Thorough this paper, we use the subscript $\mathrm{g}$ to refer to the gas phase and $\mathrm{d},k$ to refer to the $k$th dust phase, $k$ being an integer taking all the values from $1$ to $n$. In this study, we restrict ourselves to the case were dust grains do not interact with each other (in particular, they do not grow or fragment).

\subsection{Multiple fluid formalism}
\label{sec:evoleqtwo}

In a multiple fluid formalism, each phase of the mixture is treated as a fluid, with elements composed of a mesoscopic volume of particles of the given species. Those fluid elements move with their own advection velocities. Hence, with usual notations, the equations for the conservation of density, momentum and energy for the gas and the $n$ dust phases are:
\begin{eqnarray}
\frac{\partial \rhog}{\partial t} + \nabla\cdot\left ( \rhog \vg \right) & = & 0 \label{eq:mass_gas},\\
\frac{\partial \rhodk}{\partial t} + \nabla\cdot\left ( \rhodk \vdk \right) & = & 0 \label{eq:mass_dust}, \\
\rhog \left[ \frac{\partial \vg}{\partial t} + (\vg \cdot \nabla ) \vg \right]  & = & \rhog \fg +  \sk \kk (\vdk - \vg) +  \rhog \textbf{f} \label{eq:momentum_gas},\\[1em]
\rhod \left[ \frac{\partial \vdk}{\partial t} + \left(\vdk \cdot \nabla\right) \vdk \right] & = &\rhodk \fdk  -  \kk  (\vdk - \vg) + \rhodk \textbf{f}  \label{eq:momentum_dust}, \\
\frac{\partial u}{\partial t} + (\vg \cdot \nabla) u & = &   -\frac{P_{\mathrm{g}}}{\rhog} (\nabla \cdot \vg) + \sk \frac{\kk}{\rhog} (\vdk - \vg)^{2}. \label{eq:newu}
\end{eqnarray}
The different phases are coupled by drag terms, which exchange momentum and energy between the gas and the dust phases. $\kk$ denotes the drag coefficient between the gas and the $k$th dust species and has the dimension of a mass per unit volume per unit time since it defines a drag force per unit volume. It can be either a constant or a function of the differential velocities between the phases (see \citealt{LP12b} for a discussion on the different astrophysical regimes). $\fg$ and $\fdk$ denote the forces that are specific to the gas and the dust phases respectively (i.e. gas pressure gradient or viscosity, dust radiation pressure, buoyancy forces and so forth). For simplicity, we assume an ideal gas equation of state given by
\begin{equation}
P_{\rm g} = (\gamma - 1)\rho_{\rm g} u.
\end{equation}
The total dust density $\rhod$ and the dust velocity $\vd$ are defined according to
\begin{eqnarray}
\rhod & \equiv & \sk \rhodk ,  \\ 
\vd & \equiv & \frac{1}{\rhod}\dst \sk \rhodk \vdk .
\end{eqnarray}
Summing Eqs.~\ref{eq:mass_dust} over all the dust species gives the equation of conservation for the total mass of dust,
\begin{eqnarray}
\frac{\partial \rhod}{\partial t} + \nabla\cdot\left ( \rhod \vd \right) & = & 0 \label{eq:mass_dust_full}.
\end{eqnarray}
Finally, in the multiple fluid formalism, the total density of energy of the mixture is given by
\begin{equation}
e = \frac{1}{2}\rhog \vg^{2} + \sk \frac{1}{2}\rhodk \vdk^{2} + \rhog u .
\label{eq:defenergy}
\end{equation}

\subsection{One-fluid formalism}
\label{sec:evoleqone}

In the one-fluid formalism, particles of different species are treated as being part of the same continuous fluid called the mixture. The mixture's fluid elements are thus made of particles of different types that are advected with a single velocity $\vb\left(x,t \right)$. Each fluid element is constructed so that its mass is rigorously conserved, while the composition may vary since one species can replace another one. Differential velocities between the gas and the dust phases are not kinematic quantities anymore, but intrinsic properties of the fluid. This approach, developed by \citet{LP14a} for the specific case $n = 1$ we now generalise to any number of dust phases.

\subsubsection{Physical quantities}

The mixture's density $\rho$ is defined as being the total density of its constituents
\begin{equation}
\rho \equiv \rhog + \rhod  = \rhog + \sk \rhodk .
\label{eq:def_rho}
\end{equation}
The mixture's advection velocity $\vb$ is chosen to be the barycentric velocity of the different phases
\begin{equation}
\vb \equiv \dst \frac{\rhog \vg + \dst \sk \rhodk \vdk }{\rho} = \frac{\rhog \vg + \rhod \vd}{\rho} .
\label{eq:def_vb}
\end{equation}
The relative chemical composition of the mixture is expressed via the dust fractions of each species $\epsk$
\begin{equation}
\epsk \equiv \frac{\rhodk}{\rho},
\end{equation}
such that the total dust fraction is given by
\begin{equation}
\epsilon \equiv \sk \epsk = \frac{\rhod}{\rho} ,
\end{equation}
which sets the gas fraction as $\left( 1 - \epsilon \right)$ to conserve the total mass of a fluid element. This definition also ensures the following relation
\begin{equation}
\epsilon \vd = \sk \epsk \vk .
\end{equation}
The differential velocities between the $k$th dust phase and the gas are defined according to
\begin{equation}
\deltavk \equiv \vdk - \vg .
\label{eq:def_deltavk}
\end{equation}
Inverting Eqs.~\ref{eq:def_vb} and \ref{eq:def_deltavk}, the gas and dust velocities can be expressed as functions of the mixture's quantities as follows
\begin{eqnarray}
\vg & = & \vb - \sk \epsk \deltavk , \label{eq:newvg}\\
\vdk & = & \vb + \deltavk - \sk \epsk \deltavk, \label{eq:newvdk}\\
\vd & = & \vb + \frac{\left(1 - \epsilon \right)}{\epsilon} \sk \epsk \deltavk \label{eq:newvd} .
\end{eqnarray}
Introducing the total differential velocity $\deltav$ defined according to
\begin{equation}
\epsilon \deltav \equiv \sk \epsk \deltavk ,
\label{eq:def_deltav}
\end{equation}
Eqs.~\ref{eq:newvg} -- \ref{eq:newvd} can be rewritten
\begin{eqnarray}
\vg & = & \vb - \epsilon \deltav , \label{eq:newvg_bis}\\
\vdk & = & \vb + \deltavk - \epsilon \deltav, \label{eq:newvdk_bis}\\
\vd & = & \vb + \left(1 - \epsilon \right) \deltav \label{eq:newvd_bis}.
\end{eqnarray}
Eqs.~\ref{eq:newvg_bis} and \ref{eq:newvd_bis} are fully consistent with the definition of $\deltav$ in the limiting case $n = 1$. Similarly, by substituting Eqs.~\ref{eq:newvg} -- \ref{eq:newvdk} in Eq.~\ref{eq:defenergy}, the total density of energy of the mixture becomes
\begin{eqnarray}
e & = &  \frac{1}{2}\rho \vb^{2} + \frac{1}{2}\rho \left[ \sk \epsk \deltavk^{2} - \left( \epsilon \deltav \right)^{2} \right] + \left(1 - \epsilon \right)\rho u.
\label{eq:local_e}
\end{eqnarray}
The physical quantities defined above reduce to the one used for the one fluid formalism with a single dust species for the case $n = 1$.

\subsubsection{One-fluid equations}

Expressing Eqs.~\ref{eq:mass_gas} -- \ref{eq:newu} with the new physical quantities provides the system of equations describing the evolution of the mixture in the one-fluid formalism
\begin{eqnarray}
\frac{{\rm d} \rho}{{\rm d} t}& = & - \rho (\nabla\cdot\vb), \label{eq:genmass_rho} \\
\frac{{\rm d} \epsk}{{\rm d} t}& = & -\frac{1}{\rho} \nabla \cdot \left[ \rho\epsk \left( \deltavk - \epsilon \deltav \right) \right] \label{eq:gendtgevol} , \\
\frac{{\rm d} \vb}{{\rm d} t} & = & \left(1 - \epsd \right)\mathbf{f}_{\rm g} + \sk \epsk \mathbf{f}_{\mathrm{d}k} + \mathbf{f} \nonumber \\
&& - \frac{1}{\rho}\nabla\cdot \left[ \rho \sk  \epsk \deltavk \left(\deltavk -  \epsilon \deltav \right) \right] \label{eq:genmomentum_bary},\\[1.em]
\frac{{\rm d} \deltavk}{{\rm d} t}  & = &  - \dst \frac{\deltavk}{\epsk \tbk} - \dst \sum_{l} \frac{\deltavl}{\left(1 - \epsilon \right)\tbl}   \nonumber \\[0.5em]
&&+  (\fdk - {\bf f}_{\rm g}) - (\deltavk \cdot \nabla) \vb \nonumber \\
&&+ \frac{1}{2}\nabla \left[\deltavk \cdot \left(\deltavk - 2 \epsilon \deltav \right) \right], \label{eq:genmomentum_deltav} \\
\frac{{\rm d} u}{{\rm d} t}&  = &   -\frac{ P_{\mathrm{g}}}{\left(1 - \epsd \right) \rho} \nabla\cdot \vg \nonumber \\
&&+   \epsilon \deltav. \nabla u  + \sk   \frac{\deltavk^{2}}{\left(1 - \epsilon \right)\tbk}. \label{eq:newusingle}
\end{eqnarray}
where the comoving derivative refers to a particle moving with the barycentric velocity $\vb$, i.e.
\begin{equation}
\frac{{\rm d}}{{\rm d}t} \equiv \frac{\partial}{\partial t} + (\vb. \nabla) ,
\end{equation}
and the drag stopping times $\tbk$ are given by

\begin{equation}
\tbk  =   \frac{\rho}{\kk} .
\end{equation}

Eq.~\ref{eq:genmass_rho} shows that, locally, the mass of the mixture is conserved exactly. This property has been obtained by construction, using the properties of the centre of mass of a physical system (Eqs.~\ref{eq:def_rho} -- \ref{eq:def_vb}). Eq.~\ref{eq:gendtgevol} expresses the fact that although the mass of a fluid element is constant, its composition may evolve, depending on the relative dust and gas fluxes. Eq.~\ref{eq:genmomentum_bary} shows that the mixture evolves under the action of all the forces acting on its constituents, as well as a generalised anisotropic pressure term due to momentum transferred through composition modification. Differential velocities evolve under the action of both conservative and dissipative terms (Eq.~\ref{eq:genmomentum_deltav}), which both transfer energy from a dust to the gas phase (Eq.~\ref{eq:newusingle}). This system of equations reduces exactly to the one studied in \citet{LP14a} in the specific case $n = 1$.

\subsection{Conservative terms}
\label{sec:conservative}

Similar to the $n=1$ case, it is physically enlightening to derive the conservative part of Eqs.~\ref{eq:genmass_rho} from integral conservation laws and put the system in a conservative form. From a numerical point of view, it should be noted that switching from a primitive to a conservative form preserves the hyperbolic structure of the equations, as discussed in \citet{LP14a}.

\subsubsection{Conservation of mass}

The total mass of gas, of any dust species as well as the total mass dust  contained in a given volume $V$ are
\begin{eqnarray}
M_{\rm g} & \equiv &  \int_{V} \rhog \mathrm{d}V =  \int_{V} \left(1 - \epsilon \right)\rho \mathrm{d}V, \label{eq:def_mg}\\
M_{\rm dk} & \equiv &  \int_{V} \rhodk \mathrm{d}V =  \int_{V} \epsk \rho \mathrm{d}V, \label{eq:def_mdk}\\
M_{\rm d} & \equiv & \sk M_{\rm dk} =   \int_{V}  \epsilon \rho \mathrm{d}V = \int_{V} \rhod \mathrm{d}V. \label{eq:def_md} 
\end{eqnarray}
The mass conservation for every species over the volume $V$  (including the gas) can be expressed as
\begin{eqnarray}
\frac{\mathrm{d}_{\rm g} M_{\rm g}}{\mathrm{d} t} &  = & 0  , \label{eq:int_mg} \\
\frac{\mathrm{d}_{\rm dk} M_{\rm dk}}{\mathrm{d} t} &  = & 0  , \label{eq:int_mdk}
\end{eqnarray}
where $\dst \frac{\mathrm{d}_{\rm g}}{\mathrm{d} t}  = \dst  \frac{\partial}{\partial t} + \mathbf{v}_{\rm g}\cdot\nabla$ and$\dst  \frac{\mathrm{d}_{\rm dk}}{\mathrm{d} t}  =\dst  \frac{\partial}{\partial t} + \mathbf{v}_{\rm dk}\cdot\nabla$  are the comoving derivatives for the gas, the $n$ dust species and the entire dust phase respectively. Applying the transport theorem  and the divergence theorem (similarly to \citealt{LP14a}) on Eqs.~\ref{eq:int_mg} -- \ref{eq:int_mdk} gives
\begin{eqnarray}
\frac{\partial \rho \left(1 -  \epsilon\right)}{\partial t} + \nabla\cdot\left[ \rho\left( 1 - \epsilon\right)\left(\vb - \epsilon \deltav \right) \right] & = & 0 , \label{eq:cons_mg}\\
\frac{\partial \rho \epsk}{\partial t} + \nabla\cdot\left[ \rho \epsk \left( \vb + \deltavk - \epsilon \deltav \right) \right] & = & 0. \label{eq:cons_mdk} 
\end{eqnarray}
Summing Eqs.~\ref{eq:cons_mg} and the $n$ equations of Eq.~\ref{eq:cons_mdk} gives
\begin{equation}
\frac{\partial \rho}{\partial t} + \nabla\cdot\left( \rho \vb \right)  =  0 ,
\label{eq:mass_cons}
\end{equation}
which is rigorously equivalent to
\begin{equation}
\frac{\mathrm{d} M }{\mathrm{d}t} = 0 ,
\end{equation}
where $M$ is the total mass of material contained in the volume $V$. This result is not surprising since the mixture has been constructed to exploit the conservative properties of the centre of mass of the system.  Similar to the case $n = 1$, the mass of each species or phase taken individually is \textit{not} conserved since
\begin{eqnarray}
\frac{\mathrm{d} M_{\rm g}}{\mathrm{d} t} & = & \int_{S} \rho \left(1 - \epsilon \right)  \epsilon \deltav \cdot\mathbf{n} \mathrm{d}S ,\label{eq:int_mg_bary}\\
\frac{\mathrm{d} M_{\rm dk}}{\mathrm{d} t} & = & - \int_{S} \rho \epsk\left(\deltavk - \epsilon \deltav \right)\cdot\mathbf{n} \mathrm{d}S. \label{eq:int_mdk_bary} 
\end{eqnarray}
The right-hand sides of Eqs.~\ref{eq:int_mg_bary} -- \ref{eq:int_mdk_bary} represent the fluxes of mass of each species through the surface $S$ of the volume $V$. By construction, those fluxes cancel each other when summing over the different species. 

Summing only over the $n$ equations of Eq.~\ref{eq:cons_mdk} leads to conservation relations related to the evolution of the entire dust phase
\begin{equation}
\frac{\partial \rho \epsilon}{\partial t} + \nabla\cdot\left[ \rho \epsilon \vb + \rho\left(1 - \epsilon \right) \epsilon \deltav  \right]  =  0 ,\label{eq:cons_md} \\
\end{equation}
which is equivalent to
\begin{equation}
\frac{\mathrm{d} M_{\rm d}}{\mathrm{d} t}  =  - \int_{S} \rho\left(1 - \epsilon \right) \epsilon \deltav  \cdot\mathbf{n} \mathrm{d}S ,
\label{eq:int_mdk_bary2} \\
\end{equation}
or
\begin{equation}
\frac{\mathrm{d}_{\rm d} M_{\rm d}}{\mathrm{d} t}   =  0  , 
\end{equation}
where $\dst  \frac{\mathrm{d}_{\rm d}}{\mathrm{d} t}  =\dst  \frac{\partial}{\partial t} + \mathbf{v}_{\rm d}\cdot\nabla$ is the comoving derivative of the entire dust phase. It is worth noting that the terms in Eqs.~\ref{eq:cons_md} and \ref{eq:int_mdk_bary}, though here with generalised meaning, are the same as in the $n=1$ case.

\subsubsection{Conservation of momentum}

The total momentum of gas, of dust of each species and the total momentum of dust in the volume $V$ are
\begin{eqnarray}
\mathbf{P}_{\rm g} & \equiv &  \int_{V} \rhog \vg \mathrm{d}V =  \int_{V} \rho \left(1 - \epsilon \right)\left(\vb -  \epsilon \deltav \right) \mathrm{d}V, \label{eq:def_pg}\\
\mathbf{P}_{\rm dk} & \equiv &  \int_{V} \rhodk \vdk \mathrm{d}V = \int_{V} \rho \epsk \left(\vb + \deltavk  - \epsilon \deltav \right) \mathrm{d}V, \label{eq:def_pdk}\\
\mathbf{P}_{\rm d} & \equiv & \sk \mathbf{P}_{\rm dk} =   \int_{V} \left[ \rho \epsilon \vb + \rho \left(1 - \epsilon \right) \epsilon \deltav \right] \mathrm{d}V , \nonumber \\
& = &  \int_{V} \rhod \vd \mathrm{d}V. \label{eq:def_pd} 
\end{eqnarray}
Using $P$ to denote the gas pressure, the conservation of momentum for every species reads
\begin{eqnarray}
\frac{\mathrm{d}_{\rm g} \mathbf{P}_{\rm g}}{\mathrm{d} t} &  \equiv & -\int_{S} P \mathbf{n}\mathrm{d}S , \label{eq:int_pg} \\
\frac{\mathrm{d}_{\rm dk} \mathbf{P}_{\rm dk}}{\mathrm{d} t} &  \equiv & 0. \label{eq:int_pdk}
\end{eqnarray}
Eqs.~\ref{eq:int_pg} -- \ref{eq:int_pdk} therefore result in local conservation equations given by
\begin{eqnarray}
\frac{\partial \rho \left( 1 - \epsilon \right)\left(\vb  - \epsilon \deltav \right)}{\partial t} \nonumber \\ 
 + \nabla\cdot\left[ \rho \left( 1 - \epsilon\right)\left(\vb  - \epsilon \deltav \right) \left(\vb  - \epsilon \deltav \right)+ P \mathrm{\mathbf{I}}  \right]    &=& 0 , \label{eq:cons_pg} \\
\frac{\partial \rho\epsk\left(\vb + \deltavk - \epsilon \deltav \right)}{\partial t}  \nonumber \\
 + \nabla\cdot\left[ \rho \epsk \left(\vb + \deltavk - \epsilon \deltav \right) \left(\vb + \deltavk - \epsilon \deltav \right)  \right]    &=& 0 \label{eq:cons_pdk}.
\end{eqnarray}
Summing over all the phases of the mixture (including the gas) gives the local and the integral equations of conservation for the total momentum of the mixture
\begin{equation}
\frac{\partial \rho \vb}{\partial t} + \nabla\cdot\left[ \rho \vb \vb + P  \mathrm{\mathbf{I}} + \rho \sk \left[ \epsk \deltavk \left(\deltavk -  \epsilon \deltav \right) \right] \right)  =    0, 
\end{equation}
and
\begin{equation}
\dst \frac{\mathrm{d}\mathbf{P}}{\mathrm{d}t} =  -\int_{S} P \mathbf{n} \mathrm{d}S  -  \int_{S}\rho \sk \left[ \epsk \deltavk \left(\deltavk - \epsilon \deltav \right)\right] \cdot\mathbf{n} \mathrm{d}S ,
\end{equation}
where $\mathbf{P} \equiv \mathbf{P}_{\rm g} + \mathbf{P}_{\rm d}$. In contrast to the total mass, the total momentum $\mathbf{P}$ is \textit{not} conserved since the momentum fluxes transported by the mass fluxes specific to each species do not counterbalance each other. As for the special case $n = 1$, the overall contribution is equivalent to an anisotropic pressure gradient term, but the contribution arises here from the balance between two terms. Following the same argument, the total dust momentum carried at the dust velocity is not conserved either, i.e.
\begin{equation}
\sk \frac{\mathrm{d}_{\rm dk} \mathbf{P}_{\rm dk}}{\mathrm{d} t} \ne \frac{\mathrm{d}_{\rm d} \mathbf{P}_{\rm d}}{\mathrm{d} t} .
\end{equation}

\subsubsection{Conservation of energy}

The total energy for the gas phase and the $n$ dust species over the volume $V$ are given by
\begin{eqnarray}
E_{\rm g} & = &  \frac{1}{2}\int_{V} \rhog \vg^{2} \mathrm{d}V =  \frac{1}{2} \int_{V}  \left(1 - \epsilon \right)  \rho \left(\vb -  \epsilon \deltav \right)^{2} \mathrm{d}V ,\\
E_{\rm dk} & = &  \frac{1}{2}\int_{V} \rhodk \vdk^{2} \mathrm{d}V = \frac{1}{2} \int_{V} \rho \epsk \left(\vb + \deltavk  - \epsilon \deltav \right)^{2} \mathrm{d}V.
\end{eqnarray}
Conservation of energy can therefore be expressed as
\begin{eqnarray}
\frac{\mathrm{d}_{\rm g} E_{\rm g}}{\mathrm{d} t} & = & -\int_{S} P  \left(\vb - \epsilon \deltav \right)  \mathbf{n} \mathrm{d}S, \label{eq:eg_cons}\\
\frac{\mathrm{d}_{\rm d} E_{\rm dk}}{\mathrm{d} t} & = & 0 \label{eq:eq:edk_cons}.
\end{eqnarray}
Combining the two local equations of conservation induced by Eqs.~\ref{eq:eg_cons} and \ref{eq:eq:edk_cons} leads to
\begin{eqnarray}
\frac{\displaystyle \partial e }{\partial t} +  \nabla \cdot \Bigg\lbrace \left(  \frac{1}{2}\rho \vb^{2} + \frac{1}{2}\rho \left[ \sk \epsk \deltavk^{2} - \left(  \epsilon \deltav \right)^{2} \right] \right) \vb  \nonumber &&\\ 
 +\frac{\rho}{2}\left[ \sk \epsk \deltavk^{2} \left( \deltavk -  \epsilon \deltav \right) + 2 \sk \epsk \vg \cdot \left( \deltavk - \epsilon \deltav \right)\deltavk   \right] \nonumber && \\ 
 + \rho\left(1 - \epsilon \right)\left(u + P_{\rm g} \right) \vg \Bigg\rbrace  & = & 0,
\label{eq:local_et}
\end{eqnarray}
where the total energy density $e$ is given by Eq.~\ref{eq:local_e}. This expression reduces to the one found in \citet{LP14a} for the case $n = 1$.

\subsubsection{Conservation of physical quantities over the entire space}

If the volume $V$ used in the equations above represents the entire space, the surface terms of the previous integrals go to zero and
\begin{equation}
\dst \frac{\mathrm{d}M}{\mathrm{d}t} = \dst \frac{\mathrm{d}\mathbf{P}}{\mathrm{d}t} = \dst \frac{\mathrm{d} E}{\mathrm{d}t} = 0 .
\label{eq:global_cons}
\end{equation}
Eq.~\ref{eq:global_cons} provides important constraints for any conservative numerical methods. For example, these conservation relations provide the basis on which one could derive the SPH equivalent of Eqs.~\ref{eq:mass_gas} -- \ref{eq:newu} in a form which is fully conservative, implying that Eq.~\ref{eq:global_cons} is satisfied to machine precision (see \citealt{LP14b}).  

\subsection{Drag terms}

\subsubsection{Drag coefficients}
\label{sec:dragcoeff}

Various drag regimes are encountered in astrophysical systems, depending on the properties of the grains and of the gas (see e.g. \citealt{LP12b} for an exhaustive discussion). In most of the situations, linear drag regimes (i.e. constant drag coefficient) are relevant, although non linear drag regimes can be experienced by large particles in highly energetic flows. This consideration is of importance for numerical simulations since efficient implicit time stepping is easier to implement in the linear case \citep{LP12b,LP14b}. From a numerical point of view, it is also important to handle drag coefficients that are not related to any physical quantities to benchmark the algorithms efficiently. Thus, we retain quite general drag coefficients in the following, except when a particular expression is specified. 

Importantly, the drag coefficients $\kk$ involved in Eqs.~\ref{eq:momentum_dust} -- \ref{eq:newu} correspond to drag forces expressed per unit volume. $\kk$ is therefore related to the drag coefficient of a single grain $K_{1k}$ by the relation
\begin{equation}
\kk = \rho_{\mathrm{d}k}K_{1k} / m_{k} ,
\label{eq:defKk}
\end{equation}
where $m_{k}$ denotes the mass of a single grain \citep{LP12a}. Denoting $t_{k} = K_{1k} / m_{k}$ the typical drag time exerted on a single grain, Eq.~\ref{eq:defKk} can be rewritten
\begin{equation}
\tbk = \epsk^{-1} t_{k} . 
\label{eq:defKi2}
\end{equation}

It should be noted that, as a thought experiment, a dust phase $i$ can be artificially split into several dust phases (e.g. $\epsilon_{i} = \sum_{j} \epsilon_{i,j} $). This implies also that the drag coefficients of the sub-phases should be weighted accordingly, i.e. $K_{i,j} =  \epsilon_{i,j} K_{i}$. Performing this transformation onto the drag coefficients ensures that the two descriptions of the mixture are identical. This provides a particularly efficient way of benchmarking numerical codes against analytic solutions obtained for the case $n = 1$. We have used this approach in Sec.~\ref{sec:appli}.

\subsubsection{Drag matrix}
\label{sec:dragmatrix}

Eq.~\ref{eq:genmomentum_deltav} describes the exchange of momentum between the $n$ dust phases and the gas. If we restrict the evolution of the differential velocities to the contributions of the drag terms (i.e. excluding intrinsic and external forces, as well as convective terms), we obtain the following equation
\begin{equation}
\left( \frac{\partial \deltabv}{\partial t} \right)_{\mathrm{drag}} = - \Odn \deltabv ,
\label{eq:drag_ode}
\end{equation}
where $\deltabv$ denotes the vector whose components are $\deltabvi = \deltav_{i}$, and $\Odn$ is the drag matrix defined by
\begin{equation}
\mathrm{\Omega}_{n,ij} =
\begin{cases}
\frac{1}{\dst \left(1 - \epsilon \right) t_{\mathrm{b}j}} , & i \ne j ; \\
\frac{1}{\dst t_{\mathrm{b}i}} \left( \frac{1}{\dst \epsilon_{i}  } +  \frac{1}{\dst \left(1 - \epsilon \right)  }\right) ,& i = j .
\end{cases}
\end{equation}
In the case where the mixture is composed by a single dust species only, Eq~\ref{eq:drag_ode} reduces to a simple scalar differential equation (e.g. \citealt{LP14a}). 

We now examine the properties of the matrix $\Odn$ to interpret the physics contained in Eq.~\ref{eq:drag_ode}. We first note that $\Odn$ is a diagonal plus rank-one matrix, i.e. $\Odn = \mathrm{D} + \mathrm{U}$, where
\begin{eqnarray}
 \mathrm{D}_{ij} & = & \delta_{ij} \left[\epsilon_{i} t_{\mathrm{b}i} \right]^{-1} , \label{eq:def_diag}\\
\mathrm{U}_{ij} & = & u_{i}v^{\mathrm{T}}_{j} ,
\end{eqnarray}
with
\begin{eqnarray}
u_{i} & = &  1 ,\\
v_{i} & = & \left(  \left( 1 - \epsilon\right) t_{\mathrm{b}j} \right)^{-1} .
\end{eqnarray}
Using the formula $\mathrm{det}\left( \mathrm{D} + uv^{\mathrm{T}} \right) = \mathrm{det}\left(D^{-1} \right)\mathrm{det}\left(\mathrm{I} + v^{\mathrm{T}}D^{-1} u \right) $, the determinant of $\Odn$ is:
\begin{equation}
\mathrm{det}\left( \Odn \right) = \left( \prod_{k} \frac{1}{\epsilon_{k} \tbk}\right) \times \left(1 + \sum_{k} \frac{\epsilon_{k}}{\left(1 - \epsilon \right)} \right)  > 0.
\end{equation}
Thus, the matrix $\Odn$ is invertible. The analytic expression of $\Odn^{-1}$ can be obtained from the Sherman-Morrison formula for diagonal plus rank-one invertible square matrices:
\begin{equation}
\left(D + uv^{\mathcal{T}} \right)^{-1} = D^{-1} - \dst \frac{D^{-1}   uv^{\mathcal{T}}D^{-1}}{1 +  v^{\mathcal{T}}D^{-1}u   },
\label{eq:SM}
\end{equation}
which gives after simplifications
\begin{equation}
\mathrm{\Omega}^{-1}_{n,ij} = \frac{1}{\mathrm{det}\left( \Odn \right)} \times
\begin{cases}
- \frac{ \dst \sum_{k \ne j} \frac{1}{\tbk}}{\dst \left( 1 - \epsilon \right)\prod_{k \ne (i,j)} \epsilon_{k}  }  , & i > j ; \\
- \frac{ \dst \sum_{k \ne i} \frac{1}{\tbk}}{\dst \left( 1 - \epsilon \right)\prod_{k \ne (i,j)} \epsilon_{k}  } , & i < j ; \\
\left(\dst \prod_{k \ne i} \frac{1}{\tbk} \right) \times \frac{\dst 1 - \epsilon_{i}}{\dst \left(1 - \epsilon \right) \prod_{k \ne i} \epsilon_{k}}  ,& i = j ,
\end{cases}
\label{eq:omd_inv}
\end{equation}
where $\tbk$ is related to drag timescale on a single grain by Eq.~\ref{eq:defKi2}. Physically, the differential energies between the dust phases and the gas are dissipated by the drag. In particular, the following inequality
\begin{equation}
\left( \frac{\mathrm{d} \deltabv}{\mathrm{d}t} \right)_{\rm drag} = -2\deltabv \cdot  \Odn \deltabv< 0 ,
\end{equation}
has to be satisfied, implying that $\Odn$ has to be positive definite. To prove this property, we introduce the diagonal matrix $\Psi$ defined by
\begin{equation}
\mathrm{\Psi}_{ij} = 
\begin{cases}
K_{i}^{-1/2} , & i = j ; \\
0 , & i \ne j 
\end{cases}
\end{equation}
which satisfies the similarity relation
\begin{equation}
\Wdn = \Psi^{-1} \Odn \Psi, 
\label{eq:similar}
\end{equation}
where $\Wdn$ is the real symmetric matrix (therefore positive definite) defined by
\begin{equation}
\mathrm{W}_{n,ij}  = \frac{1}{\left(1 - \epsilon \right)\rho}
\begin{cases}
K_{i}\left(1 + \dst \frac{\left(1 - \epsilon \right)}{\epsilon_{i}}  \right) , & i = j .\\
\sqrt{K_{i}K_{j}} , & i \ne j .
\end{cases} 
\label{eq:def_Wn}
\end{equation}
In Appendix~\ref{app:spectrum}, we demonstrate that the spectrum formed by the positive eigenvalues $\lambda_{k}$ of $\Odn$ (or equivalently $\Wdn$) satisfies
\begin{equation}
\left( \sum_{k} \left(1 - \epsk \right)\tk \right)^{-1} < \lambda_{\rm min} \le \lambda_{k} \le \lambda_{\rm max} \le \max_{k}\left(\frac{1}{\tk} \right) + \frac{1}{\left(1 - \epsilon \right)}\sum_{k} \epsk \tk^{-1} ,
\label{eq:bound_spectrum}
\end{equation}
Physically, the quantities $t_{\mathrm{d}k} = \lambda_{k}^{-1}$ are the inverses of the $n$ physical drag timescales encountered in the problem. \textit{A priori}, those values depart from the $n$ individual stopping times obtained when the gas and a dust phase  are treated independently to the other dust phases. Those drag timescales $t_{\mathrm{d}k}$ depend both on the drag coefficients, but also on the relative densities of each phase. This generalises the case $n=1$, for which the physical processes induced by the drag are determined by the values of the drag coefficient and the dust fraction. In a multiple dust species mixture, dense grains phases provide an efficient backreaction onto the gas. On the other hand, grains behave as individual particles in dilute dust phases. They are dragged by the gas which is itself affected by the backreaction of the dense dust phases. The dynamics of the mixture induced by the drag is therefore related to the efficiency of the coupling between the gas and the different grains species, as well as to the relative densities of the different phases.

\subsubsection{Explicit timestepping criterion}

The drag terms in Eqs.~\ref{eq:genmomentum_deltav} -- \ref{eq:newusingle} are usually integrated numerically by an operator splitting method, meaning that the drag contribution is treated independently from the conservative part of the evolution equations. In a single-fluid formalism, integration schemes for drag terms are much easier to derive than in a multiple fluid formalism (e.g. \citealt{LP14b}), since all the physical quantities required are carried by the same fluid element and no interpolation over the different phases is required.

The simplest explicit solver for Eq.~\ref{eq:drag_ode} is the forward Euler scheme
\begin{equation}
\frac{\deltabv^{n+1} - \deltabv^{n} }{\Delta t} = - \Odn^{n}  \deltabv^{n} .
\label{eq:explicit}
\end{equation}
To determine the stability constraint in Eq.~\ref{eq:explicit},  we will assume that the drag coefficients are constant. In this case, the inequality in the right hand-side of Eq.~\ref{eq:bound_spectrum} provides a lower bound for the smallest drag timescale which is larger than the smallest stopping time. Therefore, it provides a Courant-Friedrichs-Levy (CFL) condition for the drag time step $\Delta t_{\rm d,one}$ that is less stringent than $\Delta t_{\rm d,multi}$, the one which would be used with a multiple fluids treatment, namely
\begin{equation}
\Delta t > \Delta t_{\rm d,one} = \left(  \max_{k}\left(\frac{1}{\epsk\tbk} \right) + \frac{1}{\left(1 - \epsilon \right)}\sum_{k} \tbk^{-1}  \right)^{-1} ,
\label{eq:CFL_drag}
\end{equation}
since
\begin{equation}
\Delta t_{\rm d,one} > \Delta t_{\rm d,multi} =  \max_{k}\left[ \frac{1}{\tbk} \left( \frac{1}{\epsk} + \frac{1}{\left( 1- \epsilon \right)} \right)  \right] .
\end{equation}
As an example, if a single dust species $i$ is submitted to a very strong drag such that $t_{\mathrm{b}i} \ll t_{\mathrm{b} k \ne i}$, Eq.~\ref{eq:CFL_drag} can be approximated by
\begin{equation}
 \Delta t_{\rm d,one} \simeq \left( \frac{1}{ \min_{k}\left(t_{\mathrm{s}k} \right) }+ \frac{1}{\left(1 - \epsilon \right)t_{\mathrm{b}i}}  \right)^{-1} .
 \label{eq:dtone_simp}
\end{equation}
Eq.~\ref{eq:dtone_simp} shows that $ \Delta t_{\rm d,one}$ results from a balance between density weighted contributions of the $n$ stopping times and the intrinsic drag time $t_{\mathrm{b}i}$ that depends only on the drag coefficient $K_{i}$.

\subsubsection{Implicit timestepping}
\label{sec:implicit}
In numerical simulations, drag stopping times that are much smaller than all the other typical times involved in the problem induce prohibitive computational costs with explicit numerical schemes. To get rid of this issue, this conditionally stable explicit scheme has to be replaced by an unconditionally stable implicit scheme. The simplest for integrating Eq.~\ref{eq:drag_ode} is the backward Euler scheme
\begin{equation}
\frac{\deltabv^{n+1} - \deltabv^{n} }{\Delta t} = - \Odn^{n+1}  \deltabv^{n+1},
\label{eq:implicit}
\end{equation}
which is equivalent to
\begin{equation}
\deltabv^{n+1} = \left(\mathrm{I} + \Odn^{n+1} \Delta t \right)^{-1}\deltabv^{n} ,
\label{eq:inv_mat}
\end{equation}
showing that the scheme's efficiency is obtained at the price of a fast and robust matrix inversion. Using Eq.~\ref{eq:similar} to transform Eq.~\ref{eq:drag_ode}, the problem can be reduced to
\begin{equation}
\deltabvh^{n+1} = \left(\mathrm{I} + \Wdn^{n+1} \Delta t \right)^{-1}\deltabvh^{n} ,
\label{eq:inv_mat_good}
\end{equation}
where $\deltabvh = \Psi^{-1} \deltabv$. The vector $\deltabvh$ is straightforward to compute from $\deltabv$ (and vice-versa) since $\Psi$ is an analytic diagonal matrix. The general inverse problem (Eq.~\ref{eq:inv_mat}) has thus been reduced to the inversion of a real symmetric matrix, for which robust and fast algorithms are known to converge (e.g. Cholesky decomposition, Gauss-Seidel iterations). Alternatively, the Sherman-Morrison formula (Eq.~\ref{eq:SM}) can be used to invert the matrix on the right-hand side of Eq.~\ref{eq:inv_mat_good} analytically if the drag coefficient is constant. The resulting expression is however useful only for situations where the number of dust phases $n$ is not too large.

\subsection{A two-dust population model}
\label{sec:twodust}

To understand how different phases of a mixture with multiple dust species interact with each other, it is instructive to consider the special case $n=2$ involving two dust phases. Here the parameter space is narrower than for an arbitrary number of dust phases, but aspects specific to multiple dust populations remain. In this case we use
$\beta \equiv t_{1} / t_{2}$ to denote the ratio of the two single-grains drag times and $\phi_{1}$ to denote the relative dust fraction, i.e.
\begin{equation}
\phi_{1} \equiv \epsilon_{1} /  \epsilon ,
\end{equation}
which implies $\epsilon_{2} = \left( 1- \phi_{1}\right)\epsilon$. Thus, the problem is symmetric with respect to the transformation $\left[ \beta \to 1 / \beta, \phi_{1}\to \left(1 - \phi_{1} \right) \right]$. The matrix $\Omega_{2}$ becomes
\begin{equation}
\Omega_{2} = \frac{\epsilon \phi_{1}}{ t_{1} \left( 1 - \epsilon \right)} \dst
\begin{pmatrix}
\dst  1 + \frac{ \left( 1 - \epsilon \right)}{\epsilon \phi_{1}}  & \dst  \frac{\beta \left(1 - \phi_{1} \right)}{\phi_{1}} \\
\dst 1 & \dst  \frac{\beta \left(1 - \phi_{1} \right)}{\phi_{1}}  \left( 1 + \frac{ \left( 1 - \epsilon \right)}{\epsilon  \left(1 - \phi_{1} \right)} \right) 
\end{pmatrix} .
\label{eq:pbtwo}
\end{equation}
The  two physical drag time scales $t_{\mathrm{d}\pm}$ are related to the eigenvalues $\lambda_{\pm}$ of the matrix $\Omega_{2}$ by the relation
\begin{equation}
t_{\mathrm{d}\pm}^{-1} = \lambda_{\pm} =  \frac{1}{2}\left( t_{\mathrm{s}1} ^{-1}+ t_{\mathrm{s}2}^{-1} \right) \left\lbrace1 \pm \sqrt{1 - Q} \right\rbrace ,
\label{eq:eigen_two}
\end{equation}
where
\begin{equation}
Q = \frac{4 \beta\left(1 - \epsilon \right)}{\left[ \left(1 - \phi_{1} \right)\left(1 - \epsilon\left(1 - \phi_{1} \right) \right) + \beta\phi_{1}\left(1 - \epsilon \phi_{1} \right)  \right]^{2}} ,
\end{equation}
and $t_{\mathrm{s} i}$ are the usual stopping times $t_{\mathrm{s}i}^{-1} = K_{i}\left(\rho_{\mathrm{g}}^{-1} + \rho_{\mathrm{d}i}^{-1} \right)$ defined for a single dust phase mixture ($Q<1$ since $\lambda_{\pm}>0$). Thus, if $Q\to 1$, we have
\begin{equation}
t_{\mathrm{d}\pm}^{-1} = \lambda_{\pm} \simeq  \frac{1}{2}\left( t_{\mathrm{s}1} ^{-1}+ t_{\mathrm{s}2}^{-1} \right) ,
\end{equation}
and the expression is dominated by the smallest stopping time. If $Q\ll 1$,
\begin{eqnarray}
t_{\mathrm{d}+}^{-1} &  = &  \lambda_{+}  \simeq  \left( t_{\mathrm{s}1} ^{-1}+ t_{\mathrm{s}2}^{-1} \right)  ,\\
t_{\mathrm{d}-}^{-1} &  = & \lambda_{-}   \simeq  Q \left( t_{\mathrm{s}1} ^{-1}+ t_{\mathrm{s}2}^{-1} \right) .
\end{eqnarray}
In this limit, $t_{\mathrm{d}-}$ is larger than the two stopping times characterising the damping processes involved when the gas interact with the dust phase separately.

\subsection{First order approximation}
\label{sec:firstorder}

In the limit where all the $n$ drag timescales $t_{\mathrm{d}k}$ are much smaller than any other typical time scale $\tau$ involved in the problem, Eq.~\ref{eq:genmomentum_deltav} can be approximated by the so-called terminal velocity approximation (see e.g. \citealt{Youdin2005,Chiang2008,Barranco2009,Lee2010,Jacquet2011,LP14a} for applications in the case $n=1$), i.e.
\begin{equation}
\deltabv = - \Odn^{-1}  \mathbf{\Delta F} ,
\label{eq:terminal_mat}
\end{equation}
where $\mathbf{\Delta F}$ is the vector whose coordinates are the differential forces between a dust phase and the gas, i.e. $\mathbf{\Delta F}_{i} = \left(\mathbf{f}_{\mathrm{d}i} - \fg \right) $. From Eq.~\ref{eq:omd_inv}, we derived the values of each differential velocity $\deltavk$ in this strong drag limit. After simplifications we find
\begin{equation}
\deltavk = \left[\left(\mathbf{f}_{\mathrm{d}k} - \fg \right) - \sl \left(\mathbf{f}_{\mathrm{d}l} - \fg \right) \epsl  \right] \tk .
\label{eq:terminal_vec}
\end{equation}
If  $\fd = 0$, Eq.~\ref{eq:terminal_vec} reduces to
\begin{equation}
\deltavk = \fg  \left( 1 - \epsilon \right)  \tk .
\label{eq:terminal_vec_simp}
\end{equation}
Moreover, if $n= 1$, and $\fg = - \frac{\nabla P}{\rhog}$, Eq.~\ref{eq:terminal_vec} reduces to 
\begin{equation}
\deltav = \frac{\nabla P}{\rhog}\ts,
\label{eq:deltav_tradi}
\end{equation}
where the stopping time $\ts$ for a single dust phase is defined by
\begin{equation}
\ts = \frac{\rhog \rhod}{K \left( \rhog + \rhod \right)} = \frac{\epsilon \left(1 - \epsilon \right)\rho}{K} .
\label{eq:def_ts}
\end{equation}
Eq.~\ref{eq:deltav_tradi} is the usual expression for the terminal velocity in the case $n = 1$ (we used the relation $\tk = \epsilon \rho / K$ obtained from Eq.~\ref{eq:defKi2}). Using Eq.~\ref{eq:terminal_vec} to expand the evolution equations to the first order in $t_{\mathrm{d}k} / \tau$, we find
\begin{eqnarray}
\frac{{\rm d} \rho}{{\rm d} t}& = & - \rho (\nabla\cdot\vb), \label{eq:genmass_rho_one} \\
\frac{{\rm d} \epsk}{{\rm d} t}& = & -\frac{1}{\rho} \nabla \cdot \left[ \rho\epsk \left( \deltavk - \epsilon \deltav \right) \right] \label{eq:gendtgevol_one} , \\
\frac{{\rm d} \vb}{{\rm d} t} & = & \left(1 - \epsd \right)\mathbf{f}_{\rm g} + \sk \epsk \mathbf{f}_{\rm d} + \mathbf{f} , \label{eq:genmomentum_bary_one} \\
\frac{{\rm d} u}{{\rm d} t}&  = &   -\frac{ P_{\mathrm{g}}}{\left(1 - \epsd \right) \rho} \nabla\cdot \vg +   \epsilon \deltav. \nabla u  , \label{eq:newusingleone} \\
\deltavk & = & \left[\left(\mathbf{f}_{\mathrm{d}k} - \fg \right) - \sl \left(\mathbf{f}_{\mathrm{d}l} - \fg \right) \epsl  \right] \tk, \label{eq:deltavk_one}
\end{eqnarray}
since all the terms of second order arising from quadratic expressions in $\deltavk$ have being neglected.

\subsection{Zeroth-order approximation}
\label{sec:zeroth}

In the limit of an infinitely strong drag regime, $\tbk = 0$ to the zeroth order of approximation in $t_{\mathrm{d}k} / \tau$. In this limit, the gas and all the dust phases are perfectly coupled, i.e. $\deltavk = 0$. The equation of evolutions for the mixture then reduce to
\begin{eqnarray}
\frac{{\rm d} \rho}{{\rm d} t}& = & - \rho (\nabla\cdot\vb), \label{eq:genmass_rho_zero} \\
\frac{{\rm d} \epsk}{{\rm d} t}& = & 0 \label{eq:gendtgevol_zero} , \\
\frac{{\rm d} \vb}{{\rm d} t} & = & \left(1 - \epsd \right)\mathbf{f}_{\rm g} + \sk \epsk \mathbf{f}_{\rm d} + \mathbf{f}  \label{eq:genmomentum_bary_zero} , \\
\frac{{\rm d} u}{{\rm d} t}&  = &   -\frac{ P_{\mathrm{g}}}{\left(1 - \epsd \right) \rho} \nabla\cdot\vb. \label{eq:newusingle_zero} 
\end{eqnarray}
These equations are similar to the one found in the zeroth order approximation with a single dust species. Physically, this means that all the phases evolve coherently as they are stuck together by the drag, following the centre of mass of the system. In particular, the dust phases move as one and the system reduces to the case $n = 1$. This implies that the mixture can be treated like as a single gas phase with a corrected sound speed $\tilde{c}_{\rm s}$
\begin{equation}
\tilde{c}_{\rm s} = \frac{c_{\rm s}}{\sqrt{1 - \epsilon}} ,
\label{eq:cstilde}
\end{equation}
where $c_{\rm s}$ is the sound speed of the gas phase \citep{LP12a}.

\subsection{Continuous dust distributions}

\subsubsection{Physical quantities}

 So far we have assumed a finite number $n$ of dust phases. This discrete description is of practical interest for numerical simulations, for which continuous dust distributions have to be sampled over a finite number of dust phases. For analytic studies, however, it may be practical to directly use the evolution equations for a continuous dust distribution. Hence, we can describe a dust distribution depending on a single continuous parameter, the grain size $s$ (i.e. all the grains of the same size are treated as belonging to the same continuous dust phase). Here $n(s)$, $m(s)$ and $\vd(s)$ denote the number density of grains per unit size, the individual mass of a grain and the velocity of a the phase made of grains of size $s$, respectively (where $m(s) = \dst \frac{4}{3}\pi \rho s^{3}$ for compact spherical grains). The dust densities and velocities are then defined according to
\begin{eqnarray}
\rhod & \equiv & \int n(s)m(s)\ds , \\
\vd & \equiv & \frac{1}{\rhod} \int n(s)m(s) \vd(s)\ds. 
\end{eqnarray}
Thus, the definition of the mixture's density, $\rho = \rhog + \rhod$, holds. An important quantity is the dust fraction per unit size $s$, defined as
\begin{equation}
\teps(s) \equiv \frac{n(s)m(s)}{\rho} ,
\end{equation}
which satisfies
\begin{equation}
\epsilon = \int \teps(s) \ds = \dst \frac{\rhod}{\rho} .
\label{eq:def_epsiloncont}
\end{equation}
The relation given by Eq.~\ref{eq:def_epsiloncont} also ensures that $\rhog = \rho\left(1 - \epsilon \right)$ and $\vb =\left(1 - \epsilon \right)\vg + \epsilon \vd$. Using $\deltav (s) \equiv \vd(s)  - \vg$ to denote the differential velocity between grains of size $s$ and the gas, one has
\begin{equation}
\int \teps (s) \deltav(s)\ds = \epsilon \deltav ,
\end{equation}
where the generalised differential velocity for continuous dust distributions is still defined as
\begin{equation}
\deltav \equiv \vd - \vg .
\end{equation}
This implies that the gas and dust velocities can be expressed in term of the one-fluid quantities as
\begin{eqnarray}
\vg & = & \vb - \epsilon \deltav , \label{eq:vg_cont} \\
\vd & = & \vb + \deltav(s) -  \epsilon \deltav. \label{eq:vd_cont}
\end{eqnarray}
Eqs.\ref{eq:vg_cont} and \ref{eq:vd_cont} are the continuous versions of Eqs.~\ref{eq:newvg} -- \ref{eq:newvdk}. Finally, the total energy of the mixture becomes:
\begin{eqnarray}
e & = &  \frac{1}{2}\rho \vb^{2} + \frac{1}{2}\rho \left[ \int \teps(s)\deltav(s)^{2} \ds - \left(\epsilon \deltav \right)^{2} \right] + \left(1 - \epsilon \right)\rho u.
\label{eq:local_econt}
\end{eqnarray}

\begin{figure*}
\begin{center}
   \includegraphics[width=\columnwidth]{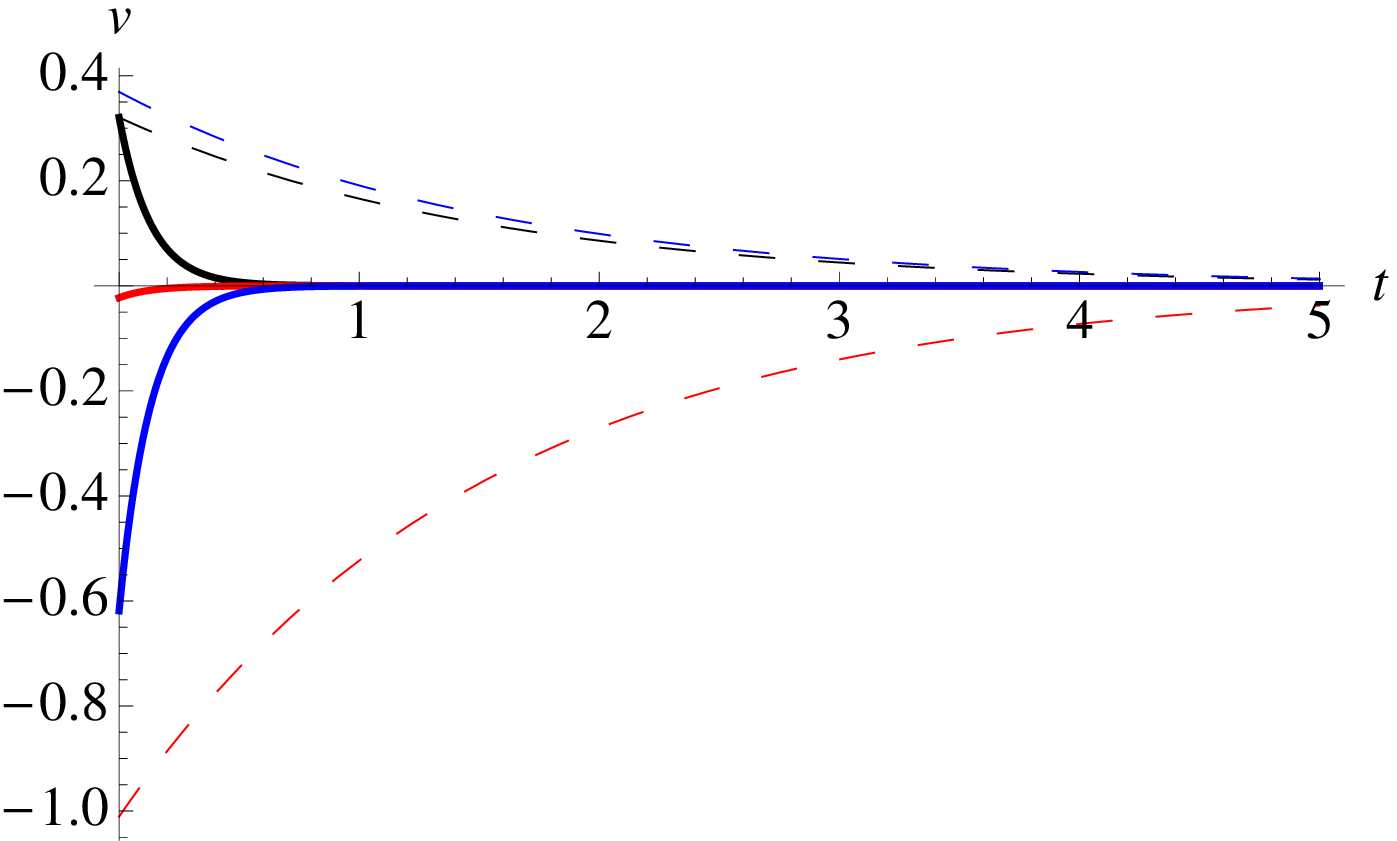}
    \hspace{0.5cm} 
       \includegraphics[width=\columnwidth]{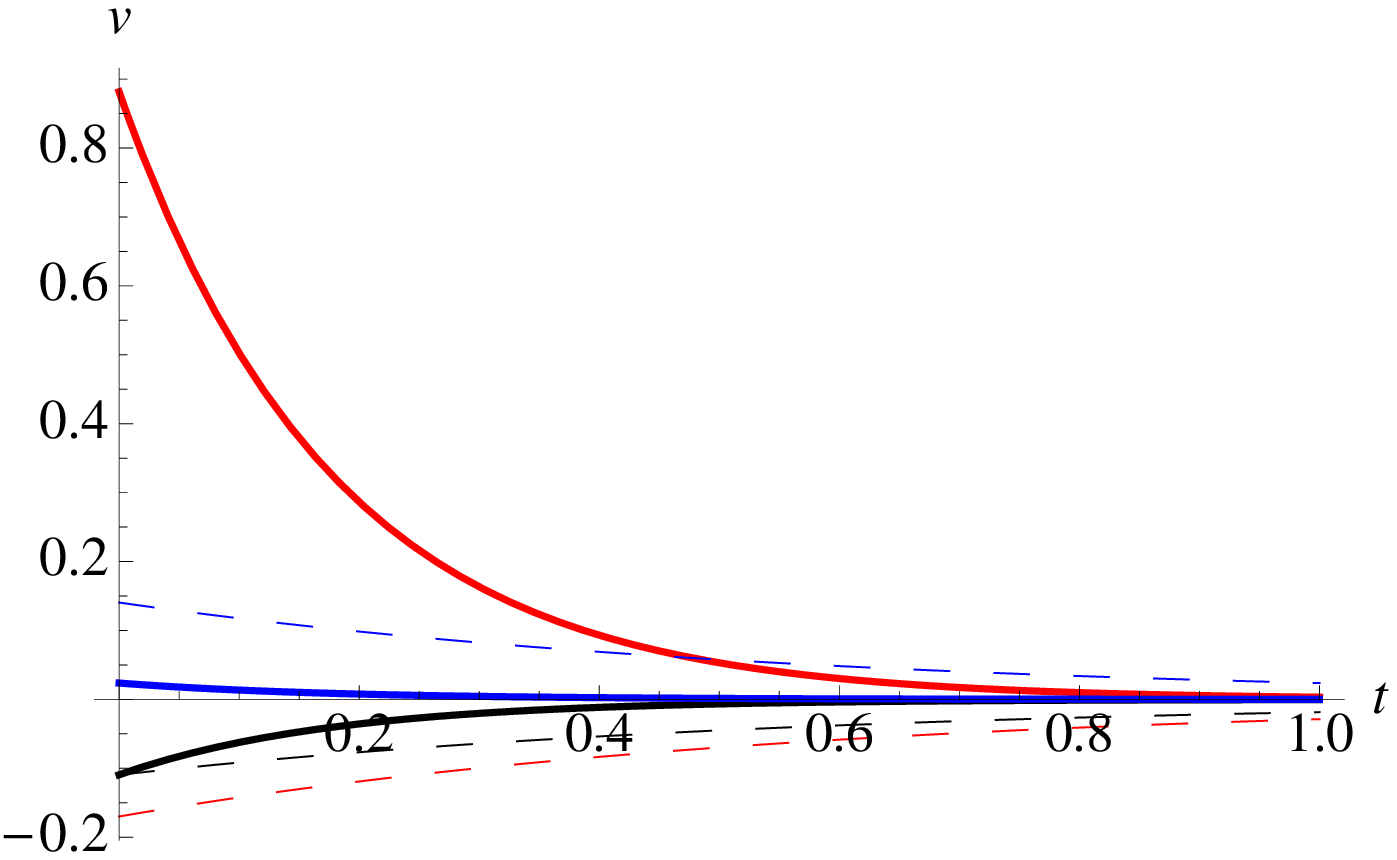}
   \caption{Evolution of gas and dust velocities towards the barycentric value in a gas + two dust phase mixture moving in opposing directions, showing the fast (thick solid lines) and the slow (thin dashed lines) eigenmodes in a linear drag regime.  Black, red and blue colours represent the gas, first and second dust phases, respectively. The parameters of the mixture are $\rho = 1$, $\epsilon = 0.5$ and $\phi_{1} = 0.5$, $t_{1} = 2$, $t_{2} = 0.2$ (left panel), $\phi_{1} = 0.1$, $t_{1} = 0.2$, $t_{2} = 1$ (right panel). Depending on the choice of parameters, the different phases evolve simultaneously or in opposition.}
   \label{fig:modeslin}
\end{center}
\end{figure*}

\subsubsection{Evolution equations}
The generalisation of Eqs.~\ref{eq:genmass_rho} -- \ref{eq:newusingle} to continuous dust distributions results in the following equations of evolution
\begin{eqnarray}
\frac{{\rm d} \rho}{{\rm d} t}& = & - \rho (\nabla\cdot\vb), \label{eq:genmass_rho_cont} \\
\frac{{\rm d} \teps(s)}{{\rm d} t}& = & -\frac{1}{\rho} \nabla \cdot \left[ \rho\teps(s) \left( \deltav(s) -\epsilon \deltav  \right) \right] \label{eq:gendtgevol_cont} , \\
\frac{{\rm d} \vb}{{\rm d} t} & = & \left(1 - \epsd \right)\mathbf{f}_{\rm g} + \int \teps(s) \mathbf{f}_{\mathrm{d}}(s)\ds + \mathbf{f} \nonumber \\
&& - \frac{1}{\rho}\nabla\cdot \left[ \rho \!\! \int \teps(s) \deltav(s) \left(\deltav(s) - \epsilon \deltav \right) \ds \right] \label{eq:genmomentum_bary_cont}, \\[1em]
\frac{{\rm d} \deltav(s)}{{\rm d} t}  & = &  - \frac{\deltav(s)}{t_{\rm s}(s)} - \int \dst \frac{\teps(s')}{\left(1 - \epsilon \right)}\frac{\deltav(s')}{t_{\rm s}(s')} \dsp   \nonumber \\[0.5em]
&&+  (\fd(s) - {\bf f}_{\rm g}) - ( \deltav(s)  \cdot \nabla) \vb \nonumber \\[0.5em]
&&+ \frac{1}{2}\nabla \left[\deltav(s) \cdot \left(\deltav(s) - 2 \epsilon \deltav \right) \right], \label{eq:genmomentum_deltav_cont} \\
\frac{{\rm d} u}{{\rm d} t}&  = &   -\frac{ P_{\mathrm{g}}}{\left(1 - \epsd \right) \rho} \nabla\cdot \vg +   \epsilon \deltav. \nabla u  + \int \dst \frac{\deltav(s)^{2}\teps(s)}{t_{\mathrm{s}}(s) \left(1 - \epsilon \right)} \ds, \label{eq:newusingle_cont}
\end{eqnarray}
where $t_{\rm s}(s)$ denotes the continuous stopping time. This is defined by
\begin{equation}
t_{\rm s}(s) =\frac{\mathrm{d}\rhod}{\mathrm{d}s}(s) = \frac{m(s)}{K_{1}(s)} ,
\end{equation}
where $K_{1}(s)$ is the drag coefficient of  a \textit{single} grain (and therefore has different dimensions to $K$, see \citealt{LP12a}). For a dust distribution characterised by single dust grain size $s_{0}$,
\begin{equation}
n(s) = \dst \frac{\rho}{m\left(s_{0} \right)} \delta \left(s - s_{0} \right) ,
\end{equation}
and
\begin{equation}
K = \int \dst K_{1}\left( s \right) s \ds = \dst \frac{\rho K_{1}\left(s_{0} \right)}{m\left(s_{0} \right)} .
\end{equation}
Eqs.~\ref{eq:genmass_rho_cont} -- \ref{eq:newusingle_cont} can also be written in a conservative form, generalising the equations derived in Sec.~\ref{sec:conservative}.

\subsubsection{Strong drag regimes}

In the limit where all the continuous dust distribution satisfies the limit of a strong drag regime, Eq.~\ref{eq:genmomentum_deltav_cont} converges to the terminal velocity approximation
\begin{equation}
\frac{\deltavk}{t_{\rm s}(s)} + \int \dst \frac{\teps(s')}{\left(1 - \epsilon \right)}\frac{\deltav(s')}{t_{\rm s}(s')} =  (\fd(s) - {\bf f}_{\rm g}) .
\label{eq:terminal_cont}  
\end{equation}
We derive the analytic solution of Eq.~\ref{eq:terminal_cont} as
\begin{equation}
\deltav(s) = \left( (\fd(s) - {\bf f}_{\rm g}) - \int  (\fd(s') - {\bf f}_{\rm g}) \teps(s') \dsp \right) t_{\rm s}(s) ,
\label{eq:terminal_contsol}  
\end{equation}
which becomes
\begin{equation}
\deltav(s) = -\fg \left( 1 -\epsilon \right) t_{\rm s}(s) ,
\end{equation}
if $\fd(s) = 0$ (a direct substitution of Eq.\ref{eq:terminal_contsol} in Eq.~\ref{eq:terminal_cont} proves the result). Eq.~\ref{eq:terminal_cont} is the continuous version of Eq.~\ref{eq:terminal_vec}. In the limit of infinitely strong drag regimes, $t_{\rm s}(s)\to 0$ and $\deltav(s) = 0$ (zeroth order approximation).

\section{Applications}
\label{sec:appli}

\subsection{\textsc{dustybox}}
\label{sec:dustybox}

The \textsc{dustybox} problem consists of gas and dust moving in opposite directions in a homogeneous, isothermal mixture, considering only the mutual drag acting between the phases. The different phases have constant uniform densities (implying $\rhog = \rho_{\mathrm{g}0}$ and $\rhodk = \rho_{\mathrm{d}k0}$, or equivalently $\rho = \rho_{0}$ and $\epsk = \epsilon_{k0}$). The initial differential velocities of the mixture as well as the gas pressure $P$ are uniform. Analytic solutions of the \textsc{dustybox} problem for different drag regimes, either linear and non-linear, are given in \citet{LP11}. Since the only forces relevant for this problem are the drag forces, the total linear momentum of the system is only exchanged between the different phases, resulting in a constant barycentric velocity ($\vb = \mathbf{v}_{0}$). As the \textsc{dustybox} problem does not involve any velocity gradient, the only relevant evolution equation is the one involving differential velocities of the mixture, which reduces to

 \begin{equation}
\frac{ \mathrm{d}\deltabv}{\mathrm{d}t} = \frac{ \partial \deltabv}{\partial t} = -\Odn \deltabv ,
\label{eq:dustybox_lin}
 \end{equation}
where $\deltabv$ is the differential velocity vector introduced in Sect.~\ref{sec:dragmatrix}. For the case of a linear drag regime, $\Odn$ has constant coefficients and the exact solution of Eq.~\ref{eq:dustybox_lin} is
\begin{equation}
\deltabv = e^{-\Odn t} \Delta \mathbf{V}_{0} .
\end{equation}
Hence, the differential velocities $\deltabv$ are progressively damped over the $n$ successive drag timescales characterising the mixture. 

\subsubsection{Results with gas and two dust phases}
We can use the two-dust phase model described in Sec.~\ref{sec:twodust} to illustrate the physics of the \textsc{dustybox} problem with multiple dust species. We set $\rho = 1$, $\epsilon = 0.5$ (so that the total mass of gas and dust are identical), $\phi_{1} = 0.5$ (the dust mass is identical in both dust phase), $t_{1} = 2$ and $t_{2} = 0.2$ (the drag is the strongest for the second phase; in practice, this would correspond to smaller grains). The eigenvalues $\lambda_{\pm}$ of the matrix $\Omega_{2}$ are given by Eq.~\ref{eq:eigen_two}. Our set of parameters gives $Q \simeq 0.3$.

We first check that the lower and upper bounds provided by Eq.~\ref{eq:bound_spectrum} are relevant. We find
\begin{equation}
0.606 <\lambda_{-} \simeq 0.659 < \lambda_{+} = 7.591 <7.750 ,
\end{equation}
showing that Eq.~\ref{eq:bound_spectrum} gives quite accurate limits for the eigenvalues of the drag matrix (we obtain similar accuracies with different parameters). The two drag timescales are $t_{1} =  \lambda_{2+}^{-1} \simeq 0.132$ and $t_{1} =  \lambda_{2+}^{-1} \simeq 1.518$. Those values differ by less than 10$\%$ from the individual stopping times  $t_{\mathrm{s}1}$ and $t_{\mathrm{s}2}$.

\begin{figure}
   \includegraphics[width=\columnwidth]{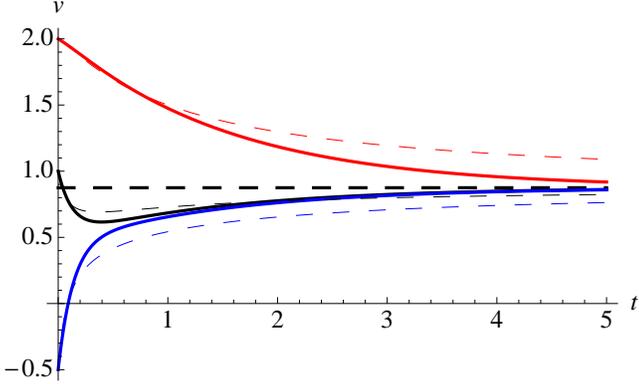}
   \caption{Comparison between the velocities obtained for the \textsc{dustybox} problems with the linear (thick solid lines) and the quadratic (thin dashed lines) drag regime in a two dust phase mixture. Parameters are similar to the ones used in Fig.~\ref{fig:modeslin}, with $v_{\rm g,0} = 1$, $v_{\rm d1,0} = 2$ and $v_{\rm d2,0} = -0.5$. Black, red and blue colours represent the gas, the first and the second dust phases respectively. The mixture's evolution is essentially identical in both cases, with the velocities converging towards the barycentric velocity $v = 1.25$ of the system (black thick dashed line). The gas velocity is not monotonic.}
   \label{fig:linquad}
\end{figure}

The left panel of Fig.~\ref{fig:modeslin} shows the velocities of the gas and the dust phases as a function of time, corresponding to the two eigenmodes of the matrix $\Omega_{2}$. The first, \emph{fast eigenmode} is the one for which the differential velocities between species is more efficiently damped. The gas velocity is in the opposite direction to both the first and the second dust species. The damping is optimal since the initial kinetic energy is mostly concentrated in the second phase, which is the most efficiently coupled to the gas. In the second, \emph{slow eigenmode} the gas and the dust species move in the opposite direction from the first dust phase, which is also the least efficiently coupled. The differential kinetic energy between the phases is thus dissipated inefficiently. The right panel of Fig.~\ref{fig:modeslin} shows that a similar behaviour is found for $t_{1} = 0.2$, $t_{2} = 1$ and $\phi_{1} = 0.1$. However, in that case the phases are coupled differently since the dust phase with the highest density is now the most poorly coupled.

\subsubsection{Quadratic vs. linear drag and non-monotonic behaviour}
Fig.~\ref{fig:linquad} compares linear and quadratic drag operators (for the quadratic case we have integrated the evolution equations numerically). The velocities of the phases were initially $v_{\rm g,0} = 1$, $v_{\rm d1,0} = 2$ and $v_{\rm d2,0} = -0.5$, and can be seen to relax towards the barycentric velocity of the system, $v =v_{0} = 1.25$. As in the $n=1$ case, the nature of the evolution is mostly independent of the drag regime. This implies that iterative numerical procedures to solve the dissipative part of the equations will work with both linear and non-linear drag regimes (as discussed above, the similar symmetric form $\Wdn$ of $\Odn$ provides the most robust structure to approximate the solution of the problem with iterative methods).

The evolution in Fig.~\ref{fig:linquad} occurs in two stages: During the first stage the gas and the second dust phase quickly stick together and form a sub-mixture composed of the gas and one dust phase. This happens in a typical time of order $t_{1}$, since the second phase possess the highest the drag coefficient and the smallest mass. In the second stage, which develops over a typical time $t_{2}$, this sub-mixture feels the drag from the first dust phase, as it possess a smaller drag coefficient and a larger density. The differential velocity between the first dust phase and the sub-mixture is then damped on a longer timescale and all the velocities of the mixture's phases converge to the barycentric velocity of the system. This example illustrates a physical property specific to multiple dust phases mixtures (i.e. $n > 1$): the evolution of the different velocities is not necessarily monotonic (which was the case for $n = 1$). In particular the gas velocity decreases and then increases under the successive actions of the second and the first dust species, respectively (see black lines in Fig.~\ref{fig:linquad}).

 We have also solved the \textsc{dustybox} problem for the $n=3$ case, finding results similar to those discussed above.

\subsection{\textsc{dustywave}}
\subsubsection{General case}

\begin{figure*}
\begin{center}
   \includegraphics[width=\columnwidth]{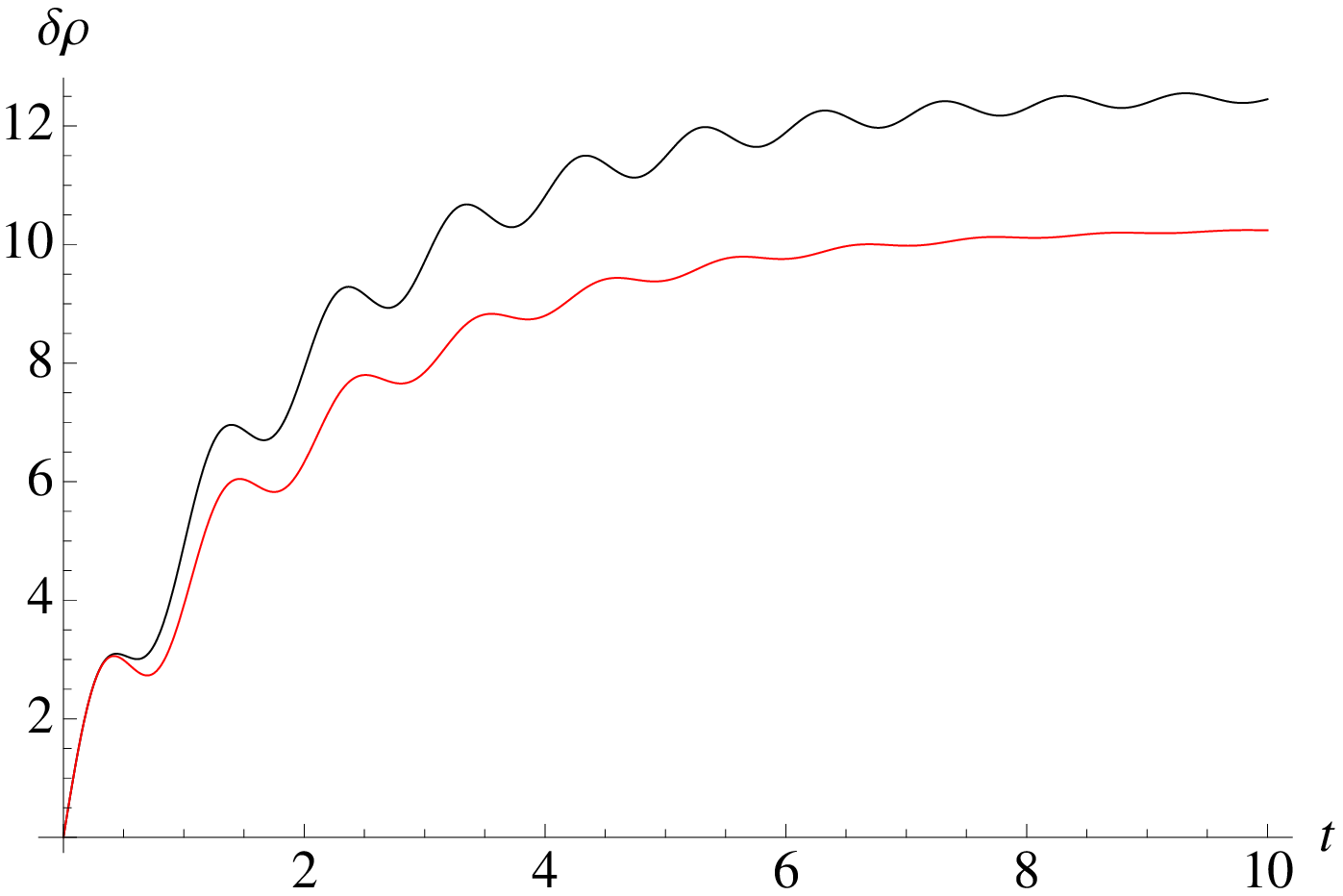}
   \hspace{0.5cm}
   \includegraphics[width=\columnwidth]{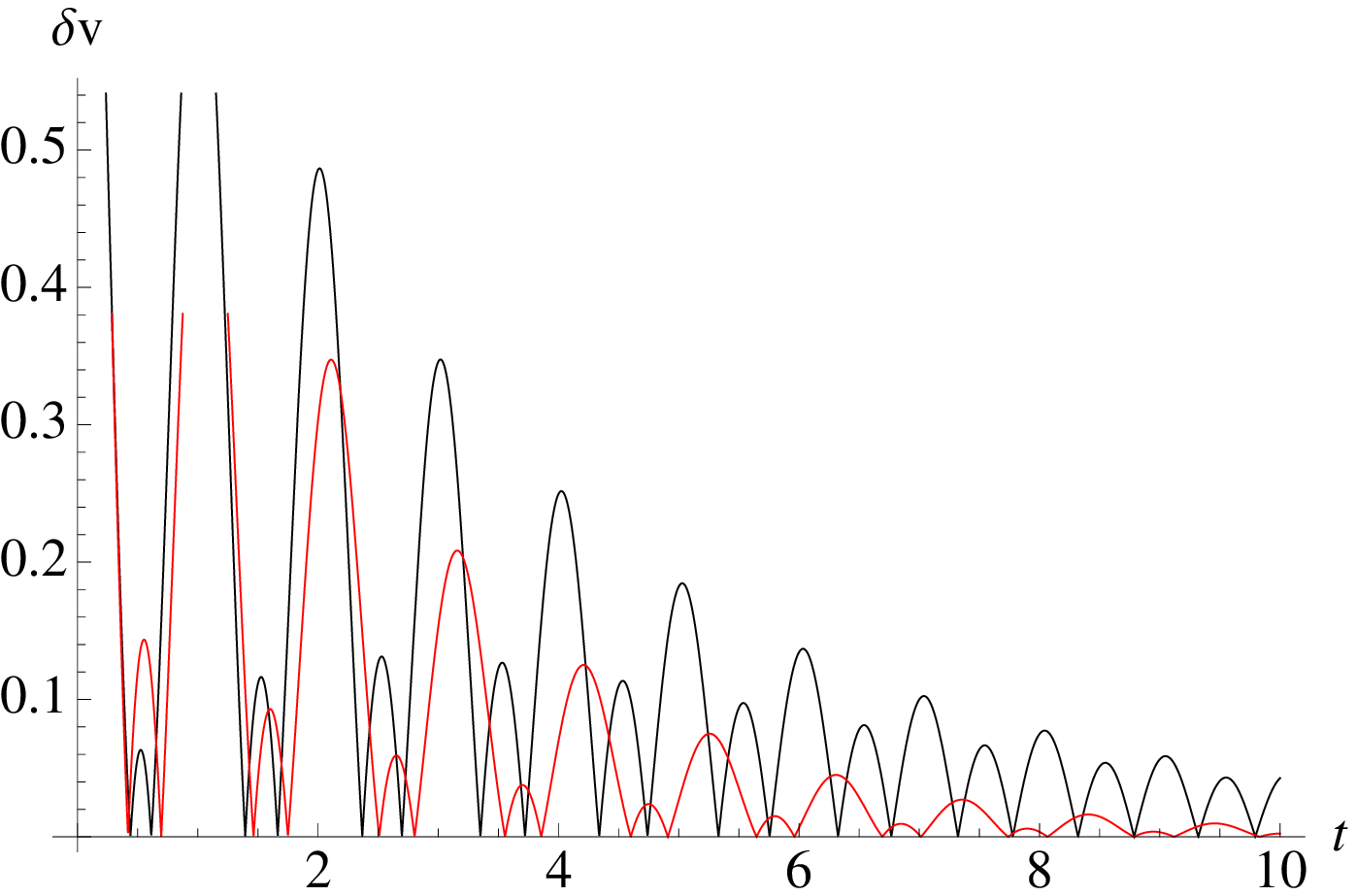} \\
   \vspace{0.5cm}     
     \includegraphics[width=\columnwidth]{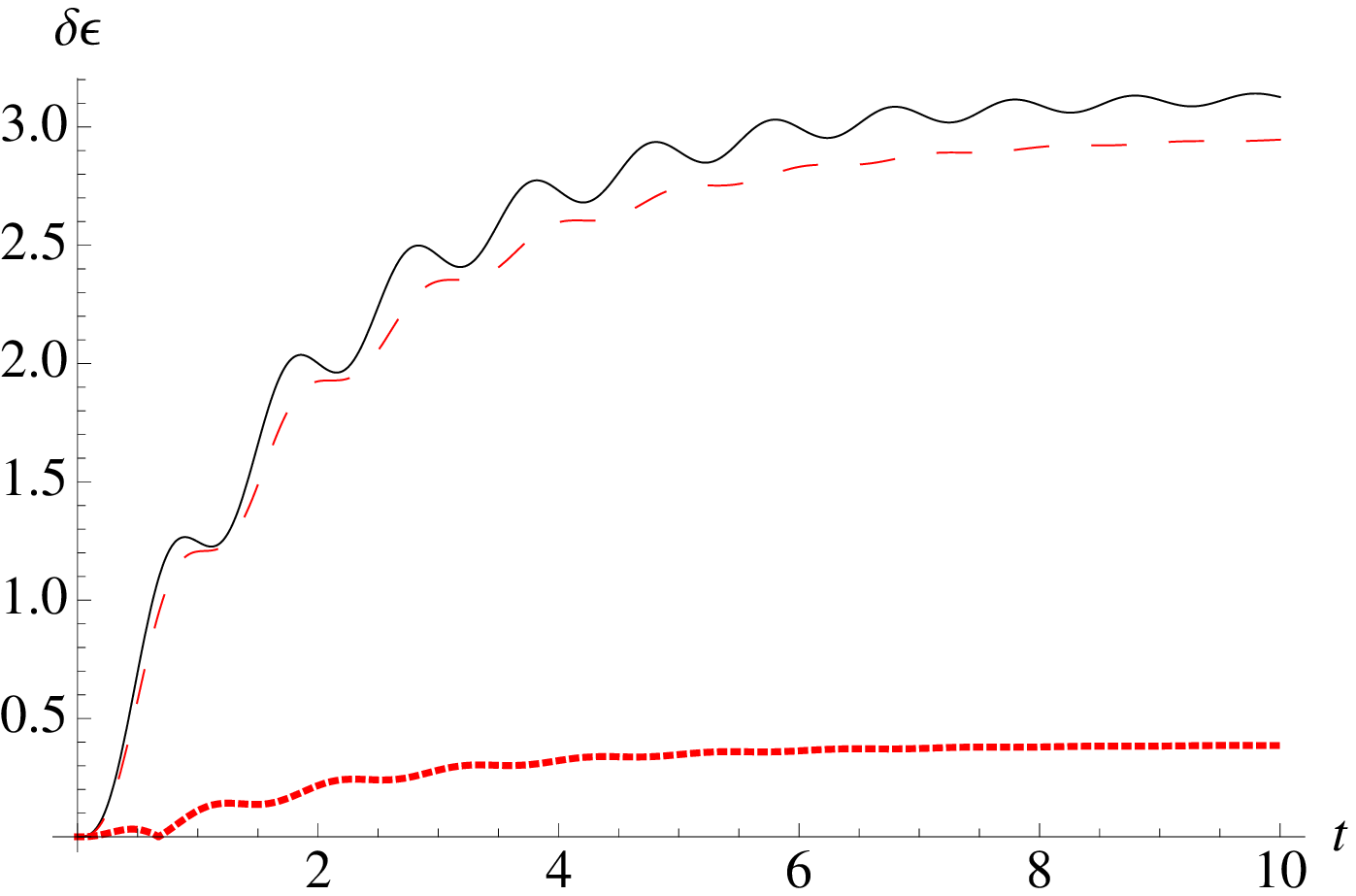}
   \hspace{0.5cm}
   \includegraphics[width=\columnwidth]{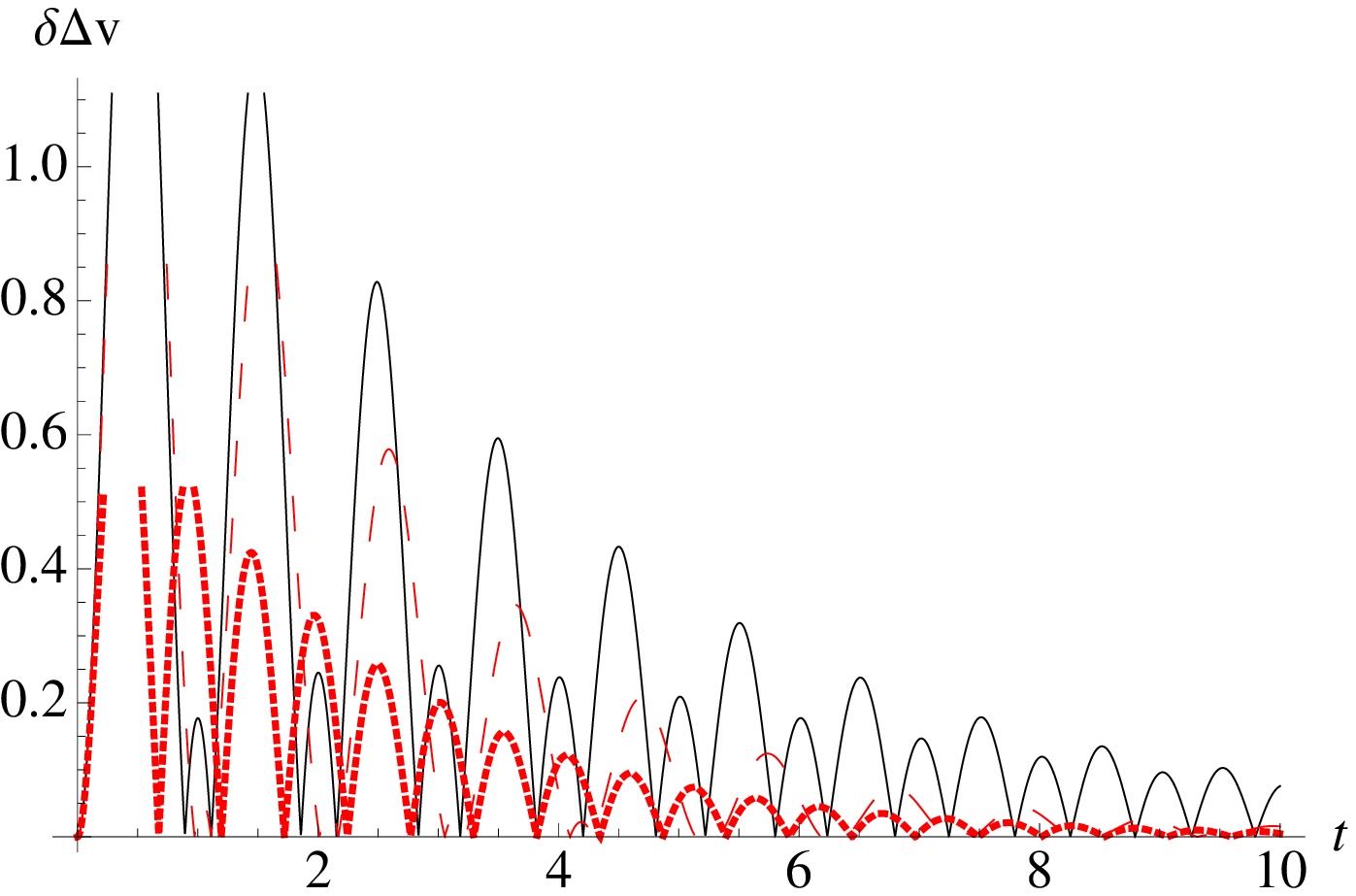}  
   \caption{Comparison of the evolution of the amplitude of the perturbations in the \textsc{dustywave} problem with one (black) and two (red) dust phases. For the $\delta \epsilon$ and the $\delta \deltav$ panels, dashed and dotted lines represent perturbations related to the first and the second dust phases, respectively. The parameters used for the background equilibrium are $c_{\rm s} = 1$, $\rho_{0} = 2$, $\epsilon_{0} = 0.5$, $t_{1} = 2$, $k = 2 \pi$ and $\phi_{1} = 0.8$, $t_{2} = 0.2$. No major differences are found between the two cases --- with the evolution of each perturbation being similar. The amplitudes of the perturbations have been renormalised to the initial velocities of the phases.}
\label{fig:dustywave}
\end{center}
\end{figure*}

The \textsc{dustywave} problem consists of the propagation of a linear acoustic wave in a dust and gas mixture, with the different phases interacting via linear drag terms. The analytic solution for the \textsc{dustywave} problem in the special case $n=1$ is provided in \citet{LP11}. Here, we generalise the problem for an arbitrary number of dust phases. Linearising the evolution equations for the mixture around the equilibrium solution $\rho = \rho_{0}$, $\epsilon = \epsilon_{0}$, $v = \Delta v  = 0$ gives
\begin{eqnarray}
\frac{\partial  \delta \rho}{\partial t} & = & - \rho_{0} \frac{\partial \delta v}{\partial x} , \label{eq:dustywave1} \\ 
\rho_{0} \frac{\partial \delta  v}{\partial t} & = & -c_{\rm s}^{2} \left[ \left(1 - \epsilon_{0} \right) \frac{\partial \delta \rho}{\partial x} - \rho_{0} \frac{\partial \delta \epsilon}{\partial x}  \right] , \label{eq:dustywave2} \\ 
\rho \frac{\partial \delta  \epsk}{\partial t} & = & -\frac{\partial}{\partial x} \left( \rho_{0}  \epsilon_{k0}\left[ - \delta \Delta v_{k} - \sum_{l} \epsilon_{l0} \delta\Delta v_{l}  \right]\right) ,  \label{eq:dustywave3} \\
 \frac{\partial \delta  \Delta v_{k}}{\partial t}  & = & -\frac{\delta \Delta v_{k}}{t_{k0}} - \sum_{l} \frac{\epsilon_{l0} }{\left( 1 - \epsilon_{0}\right)t_{l0}}\delta \Delta v_{l}  \nonumber \\
&& + \frac{\cs^{2}}{\left( 1 - \epsilon_{0}\right)\rho_{0}}  \left[ \left(1 - \epsilon_{0} \right) \frac{\partial \delta \rho}{\partial x} - \rho_{0} \frac{\partial \delta \epsilon}{\partial x}  \right]. \label{eq:dustywave4}
\end{eqnarray}
where an isothermal equation of state $\delta P  = cs^{2} \left[ \left(1 - \epsilon_{0} \right)\delta \rho - \rho_{0}\delta \epsilon \right]$ and the relation $\delta \epsilon = \sum_{k} \delta \epsilon_{k}$ have been used. To first order, the individual fluctuations of the dust fractions $\delta \epsilon_{k}$ are not involved in Eqs.~\ref{eq:dustywave1} -- \ref{eq:dustywave4} and only the terms in $\delta \epsilon$ coming from the gas pressure are relevant. The dispersion relation related to those $2n+2$ equations is a polynomial of order $2n+1$ in $\omega$ and cannot be factored easily.

We illustrate the physics of the \textsc{dustywave} problem with multiple dust species with the two dust phase mixture model described in Sect.~\ref{sec:twodust}. We assumed perturbations of the form $\delta A = \delta \tilde{A}\left( t \right) e^{ikx}$ in Eqs.~\ref{eq:dustywave1} --\ref{eq:dustywave4}, and solved the resulting system of ordinary differential equations numerically. Absolute values of the resulting complex amplitudes may then be plotted and compared to those in a mixture with a single dust phase. Fig.~\ref{fig:dustywave} shows the evolution of the real amplitudes of the perturbations in the case $n = 1$ ($c_{\rm s} = 1$, $\rho_{0} = 2$, $\epsilon_{0} = 0.5$, $t_{1} = 2$, $k = 2 \pi$ in code units) and $n=2$ ($\phi_{1} = 0.8$, $t_{2} = 0.2$). These parameters are identical to the those used in the \textsc{dustybox} problem in Sect.~\ref{sec:dustybox}. Initially, $\delta v_{\rm g} = \delta v_{\mathrm{d}1} = \delta v_{\mathrm{d}1}$ and $\delta \rho_{\rm g} = \delta \rho_{\mathrm{d}1} = \delta \rho_{\mathrm{d}1} = 0$.

 We find that the evolution of the perturbations in the $n=2$ case is similar to the $n = 1$ case. After a transient regime during which the drag terms damp the differential velocities between the dust and the gas phases, the mixture stays at rest. The entire kinetic energy of the mixture has been progressively damped by the drag, since the gas pressure maintains a non-zero differential velocity between the gas and the dust phases by propagating a perturbation at the gas sound speed $c_{\rm s}$. After several drag times, periodic density fluctuations remain in the dust phases as remnants of the sound waves dissipated by the gas drag. The asymptotic values obtained for $\delta \rho$ and $\delta \epsilon$ balance each other to give $\delta \rhog = 0$, since the energy powering the acoustic wave is entirely dissipated. We have also studied the $n = 3$ case, and found similar results. 

\subsubsection{Terminal velocity approximation}

The limiting behaviour of the \textsc{dustywave} problem in a drag dominated regime is given in \citetalias{LP12a} for the case $n = 1$. For any number of dust phases, an analytic solution can be derived using the generalised terminal velocity approximation given in Sec.~\ref{sec:firstorder}. Substituting Eq.~\ref{eq:terminal_vec_simp} into the evolution equations gives
\begin{eqnarray}
\frac{\mathrm{d}\rho}{\mathrm{d}t} & = & - \rho \frac{\partial v}{\partial x} , \label{eq:dustywave_eq1_tv} \\ 
\rho \frac{\mathrm{d}v}{\mathrm{d}t} & = & -\frac{\partial P}{\partial x} , \label{eq:dustywave_eq2_tv} \\ 
\rho \frac{\mathrm{d} \epsk}{\mathrm{d}t} & = - & \frac{\partial}{\partial x} \left( \epsk \frac{\partial P}{\partial x} \left[ \tk - \sl \epsl t_{l} \right]   \right). \label{eq:dustywave_eq3_tv} 
\end{eqnarray}
A linear expansion of Eqs.~\ref{eq:dustywave_eq1_tv} -- \ref{eq:dustywave_eq3_tv} gives
\begin{eqnarray}
\frac{\partial \delta \rho}{\partial t} & = & - \rhoz \frac{\partial \delta v}{\partial x} , \label{eq:dustywave_eq1_pert} \\ 
\rhoz \frac{\partial \delta \vb}{\partial t} & = & - cs^{2} \frac{\partial}{\partial x} \left( \left(1 - \epsz \right)\delta \rho - \rhoz \delta \epsilon \right) , \label{eq:dustywave_eq2_pert} \\ 
\rhoz \frac{\partial \delta \epsk}{\partial t} & =  &  - cs^{2}  \epskz \left[ t_{k0} - \sl \epslz t_{l0} \right] \times \nonumber \\
&& \frac{\partial^{2}}{\partial x^{2}} \left( -\epsz \delta\rho +  \rhoz \delta \epsilon \right) , \label{eq:dustywave_eq3_pert} 
\end{eqnarray}
and after summing over the indices $k$ in the $n$ equations of Eq.~\ref{eq:dustywave_eq3_pert}, we obtain
\begin{equation}
\rhoz \frac{\partial \epsilon}{\partial t} = -\cs^{2} \epsz \tse \frac{\partial^{2}}{\partial x^{2}} \left( -\epsz \delta\rho +  \rhoz \delta \epsilon \right) \label{eq:dustywave_eq4_pert} ,
\end{equation}
where $\tse$ denotes the effective stopping time of the mixture which is given by:
\begin{equation}
\tse = \frac{\left(1 - \epsz \right)}{\epsz} \sk \epskz t_{k0}  .
\label{eq;def_tse}
\end{equation}
In the case of a single dust species, Eq.~\ref{eq;def_tse} reduces to $\tse = \epsilon\left(1 - \epsilon \right) t_{\mathrm{b}} = \ts$, the usual stopping time.

Remarkably, the system formed by Eqs.~\ref{eq:dustywave_eq1_pert} -- \ref{eq:dustywave_eq2_pert} and \ref{eq:dustywave_eq4_pert} is equivalent to the system found for the case $n = 1$, simply with $\ts$ replaced by $\tse$ \citep{LP14a}. We deduce from this analogy, by extrapolating the result from the analytic solution derived in \citetalias{LP12a} for $n = 1$, that monochromatic plane waves solutions of the \textsc{dustywave} problem therefore satisfy the dispersion relation
\begin{equation}
\omega^{2} = k^{2} \cs^{2} \left[ \left(1 - \epsz \right) - i \omega \epsz \tse \right].
\label{eq:disp_rel}
\end{equation}
\subsubsection{Resolution criterion at high drag}
 Importantly, as shown in \citet{LP12a} for the case $n = 1$, Eq.~\ref{eq:disp_rel} sets the spatial resolution criterion required when simulating  strong drag regimes in a dust and gas mixture with a multiple fluid algorithm. For an arbitrary number of dust species, this criterion can be generalised to give
\begin{equation}
\Delta \lesssim \cs \tse ,
\end{equation}
where $\Delta$ is the resolution length of the simulation ($\Delta \simeq h$, the smoothing length, in SPH simulations). However, if the evolution of the gas and the dust phases are computed with a numerical method based on the one-fluid formalism, this spatial criterion resolution is irrelevant, since the mixture's differential velocities are intrinsic quantities that are advected with the fluid \citep{LP12a,LP14a,LP14b} rather than representing a physical separation of resolution elements.

\subsubsection{Drag timescales with continuous dust distributions}
For continuous dust distributions, the same reasoning can be performed, leading to
\begin{equation}
\tse = \frac{\left(1 - \epsz \right)}{\epsz} \int \teps (s) t_{\mathrm{s}}(s) \ds  .
\label{eq;def_tse_cont}
\end{equation}
Eq.~\ref{eq;def_tse_cont} shows that if the density of dust fraction $\teps (s)$ and the drag timescales of small grains go like $\teps \propto s^{a}$ and $t_{\mathrm{s}} \propto s^{b}$, the numerical value of $\tse$ will be dominated by the large or small grains depending on whether the quantity $a+b+1$ takes positive or negative values, respectively. As an example, consider spherical compact grains with a size distribution typical of the ISM, $n(s) \propto s^{-3.5}$, $a = -0.5$ and grains submitted to the Epstein drag regime, for which $b = 1$. In this case we have $a+b+1 = 1.5$. This tends to indicate that within the dust population which satisfies the terminal velocity approximation, the contribution from large grains dominates over the integral summation in Eq.~\ref{eq;def_tse_cont}. This implies that the spatial resolution criterion would be less stringent than if the integral were dominated by the contribution of the small grains. However, the caveat of this simple reasoning is that $n(s)$ is not a free parameter of the problem, since it should evolve according to Eq.~\ref{eq:gendtgevol} (as well as grain growth, which is neglected here).

Replacing $\ts$ by $\tse$ is not a suitable recipe for every physical problem. In the \textsc{dustywave} problem, this result arises because: 1) the problem is linearised, 2) we are in the limit of strong drag, and 3) the pressure gradient at equilibrium is zero, implying that the perturbations in the dust fractions play a role via $\delta \epsilon$ only, and not via the individual values of $\delta \epsk$. This last point allows one to sum over the indices $k$ in order to reduce the problem to the propagation of an acoustic wave in a mixture with a single dust phase.

\subsection{\textsc{dustyshock}}

The \textsc{dustyshock} problem consists of the propagation of a 1D shock in a dust and gas mixture. As shown in \citet{Miura1982} and \citet{LP12a,LP14b}, the shock evolution is divided into two phases. First, the differential velocities between the gas and the dust are damped. In the case of a mixture with multiple dust phases, this transient regime occurs during the $n$ physical drag timescales of the problem, which are the inverses of the eigenvalues of the drag matrix $\Odn$. Then, the mixture reaches a stationary regime, where the shock propagates as in a pure gas phase, but at the modified sound speed $\tilde{c}_{\rm s}$ given by Eq.~\ref{eq:cstilde}. $\tilde{c}_{\rm s}$ is the same modified sound speed as for the case $n = 1$. Indeed, $\tilde{c}_{\rm s}$ is related to the behaviour of the mixture in the limiting regime of an infinite drag, which does not depend on the number of dust phases (see Sect.~\ref{sec:zeroth}). 

In the limiting case of strong drag regimes, similar resolution issues arise for the \textsc{dustyshock} problem as for \textsc{dustywave} problem when treating the system with a multiple fluid formalism \citep{LP12a}. This issue can be fixed by using the single-fluid formalism developed in this paper, exactly as in the one dust species case \citep{LP14a,LP14b}. In the case of a multiple dust population, using the one-fluid formalism is even more valuable since it avoids the need for high resolution everywhere merely because a small fraction of strongly coupled dust grains are present.

\subsection{Radial migration in discs}
\label{sec:NSH86}

\subsubsection{Analytic solution}
The radial-drift of single sized dust grains in protoplanetary discs is a well studied problem (e.g. \citealt{Weidendust1977,NSH86,YS2002,Laibe2012,Laibe2014a,Laibe2014b}). In a two dimensional $(x,y)$ shearing box rotating at an angular velocity $\Omega$ \citep{Goldreich1965}, the analytic solution for the problem reads:
\begin{eqnarray}
v_{x} & = & 0 ,\label{eq:NSH86_vx} \\
v_{y} & = & -\frac{3}{2} \Omega x + \frac{1}{2 \rhoz \Omega} \partial_{r}P  ,\label{eq:NSH86_vy} \\
\deltav_{x} & = & \frac{\partial_{r}P}{\rhoz \left(1 - \epsz \right)} \frac{\ts}{\left(1 + \Omega^{2}\ts^{2} \right)} ,\label{eq:NSH86_deltavx} \\
\deltav_{y} & = & - \frac{\partial_{r}P}{\rhoz \left(1 - \epsz \right)} \frac{\Omega \ts^{2}}{2\left(1 + \Omega^{2}\ts^{2} \right)} ,\label{eq:NSH86_deltavy}
\end{eqnarray}
where the large scale pressure gradient $\partial_{r}P$ is a constant over the size of the box. This large scale pressure gradient enforces a differential velocity between the gas and the dust which is damped by the drag. As a result, angular momentum is transferred from the dust (which therefore migrates inwards) to the gas (which migrates outwards). \citet{YS2002,Laibe2012,PL2014} have shown that grains can pile-up as they reach the inner regions of the disc provided that the drag intensity increases enough to balance the increased migration efficiency from the increasing radial pressure gradient. This result has been extended to the case of growing grains \citep{Laibe2014a,Laibe2014b,PL2014}, showing that a significant fraction of the classical T-Tauri Star discs should retain their particles during the initial stages of planet formation.

The analytic solution for the problem of the radial evolution of a multiple grain sizes distribution is derived in Appendix~A of \citet{Bai2010}. We show here how to rederive it from the one-fluid formalism. We first note that
\begin{eqnarray}
\mathbf{f} &  = & 3 \Omega^{2} x \mathbf{u}_{x} ,\\
\fg &  = & -  \frac{\partial_{r}P}{\rhoz \left(1 - \epsz \right)} \mathbf{u}_{x} - 2 \mathbf{\Omega} \times \vg , \\
\fdk & = &  - 2 \mathbf{\Omega} \times \vdk ,
\end{eqnarray}
so that
\begin{eqnarray}
\frac{\rhog \fg + \sk \rhodk \fdk }{\rho}& = & - \frac{\partial_{r}P}{\rhoz } \mathbf{u}_{x} - 2 \mathbf{\Omega} \times \vb  ,\\
\fdk - \fg & = &  \frac{\partial_{r}P}{\rhoz \left(1 - \epsz \right)} - 2 \mathbf{\Omega} \times \deltavk,
\end{eqnarray}
where $\mathbf{f}$ is the usual expression for the balance between the gravity from the central star and the centrifugal force in a Keplerian potential. Importantly the forces specific to each species contain the Coriolis terms, since they depend on the intrinsic velocity of each phase. 
\begin{figure}
   \includegraphics[width=\columnwidth]{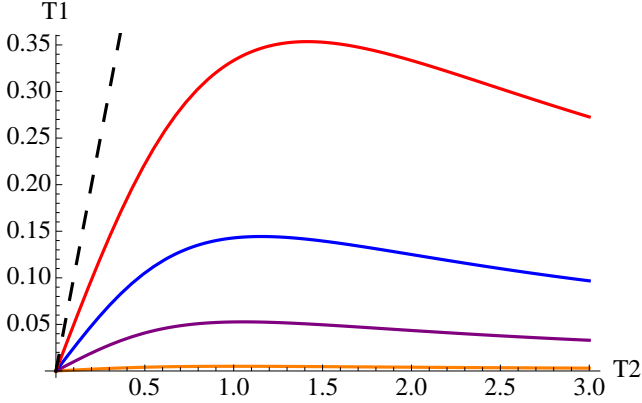}
   \caption{The orange, purple, blue and red curves give the values of the function $g_{\epsilon}\left(T_{1},T_{2} \right)$ for increasing values of the dust fraction, i.e. $\epsilon = 0.01$, $0.1$, $0.25$ and $0.5$ respectively. Below those curves, the grains of the first dust phase migrate outwards in the individual grains limit $\left( \phi_{1} = 0\right)$. This process can happen only if the grains of the other species are larger, since the curves are below the line $T_{1} = T_{2}$ (black dashed line). The maximum possible size for outwardly migrating grains (given by the maximum of the curves $g_{\epsilon}$) is an increasing function of the dust fraction $\epsilon$.}
   \label{fig:feps}
\end{figure}
We now look for stationary solutions consisting of a homogeneous perturbation superimposed on a constant shear (Eqs.~\ref{eq:genmass_rho} -- \ref{eq:gendtgevol} imply $\rho =\rhoz$ and $\epsk = \epsilon_{k0}$). The scalar equations in $v_{x}, v_{y}$ are therefore
\begin{eqnarray}
-2 \Omega v_{y} + v_{x} \frac{\partial v_{x}}{\partial x} + v_{y} \frac{\partial v_{x}}{\partial y}  & = & 3 \Omega^{2}x - \frac{\partial_{r}P}{\rhoz}, \label{eq:NSH86_multi1}\\
 2 \Omega v_{y} + v_{x} \frac{\partial v_{y}}{\partial x} + v_{y} \frac{\partial v_{y}}{\partial y} & = & 0 \label{eq:NSH86_multi2},
\end{eqnarray}
whose solution is identical to the single dust species case and is given by Eqs.~\ref{eq:NSH86_vx} -- \ref{eq:NSH86_vy}. Writing the $2n$ equations for the quantities $\deltavkx$, $\deltavky$ in a matrix form, we have
\begin{equation}
\begin{pmatrix}
\Odn & -2 \Omega \mathrm{I}_{n} \\
\frac{1}{2} \Omega \mathrm{I}_{n} & \Odn
\end{pmatrix}
\deltabvt = \frac{\partial_{r}P}{\rhoz \left(1 - \epsz \right)} 
\begin{pmatrix}
\mathrm{1}_{n,1} \\
\mathrm{0}_{n,1}
\end{pmatrix} ,
\label{eq:deltav_matrix}
\end{equation}
where $\deltabvti = \deltavix$ if $i \le n$, $\deltabvti = \deltaviy$ if $i > n$ and $\mathrm{1}_{n,1}$ is the column vector of dimension $n$ containing only the value unity (and similarly $\mathrm{0}_{n,1}$ contains zeros).
Since $\Odn$ is positive definite, $\mathrm{det}\left( \Odn^{2} + \Omega^{2}  \mathrm{I}_{n}  \right) > 0$ and the matrix is invertible. Using the identity
\begin{equation}
\Odn \left( \Odn^{2} + \Omega^{2}  \mathrm{I}_{n}  \right)^{-1} = \left( \Odn^{2} + \Omega^{2}  \mathrm{I}_{n}  \right)^{-1} \Odn  ,
\end{equation}
Eq.~\ref{eq:deltav_matrix} can be inverted by blocks, giving the solutions for the quantities $\deltavkx$ and $\deltavky$ as
\begin{equation}
\dst \begin{pmatrix}
\deltavkx \\
\deltavky
\end{pmatrix} 
 = \frac{\partial_{r}P}{\rhoz \left(1 - \epsz \right)} 
\begin{pmatrix}
\dst \sum_{j} \left[ \Odn \left( \Odn^{2} + \Omega^{2}  \mathrm{I}_{n}  \right)^{-1}\right]_{kj}  \\
\dst - \frac{\Omega}{2} \sum_{j} \left( \Odn^{2} + \Omega^{2}  \mathrm{I}_{n}  \right)^{-1}_{kj}
\end{pmatrix} .
\label{eq:NSHmulti_sol}
\end{equation}
It is straightforward to see that in the case of a single dust species, Eq.~\ref{eq:NSHmulti_sol} reduces to Eqs.~\ref{eq:NSH86_deltavx} -- \ref{eq:NSH86_deltavy}. We have not, however, been able to invert the matrix $ \Odn^{2} + \Omega^{2}  \mathrm{I}_{n} $ of Eq.~\ref{eq:NSHmulti_sol} analytically in an elegant way. $ \Odn^{2} + \Omega^{2}  \mathrm{I}_{n}$ is however similar to a positive definite matrix and can easily be inverted numerically. 

\subsubsection{Migration with two dust species}
 \begin{figure*}
\begin{center}
   \includegraphics[width=\columnwidth]{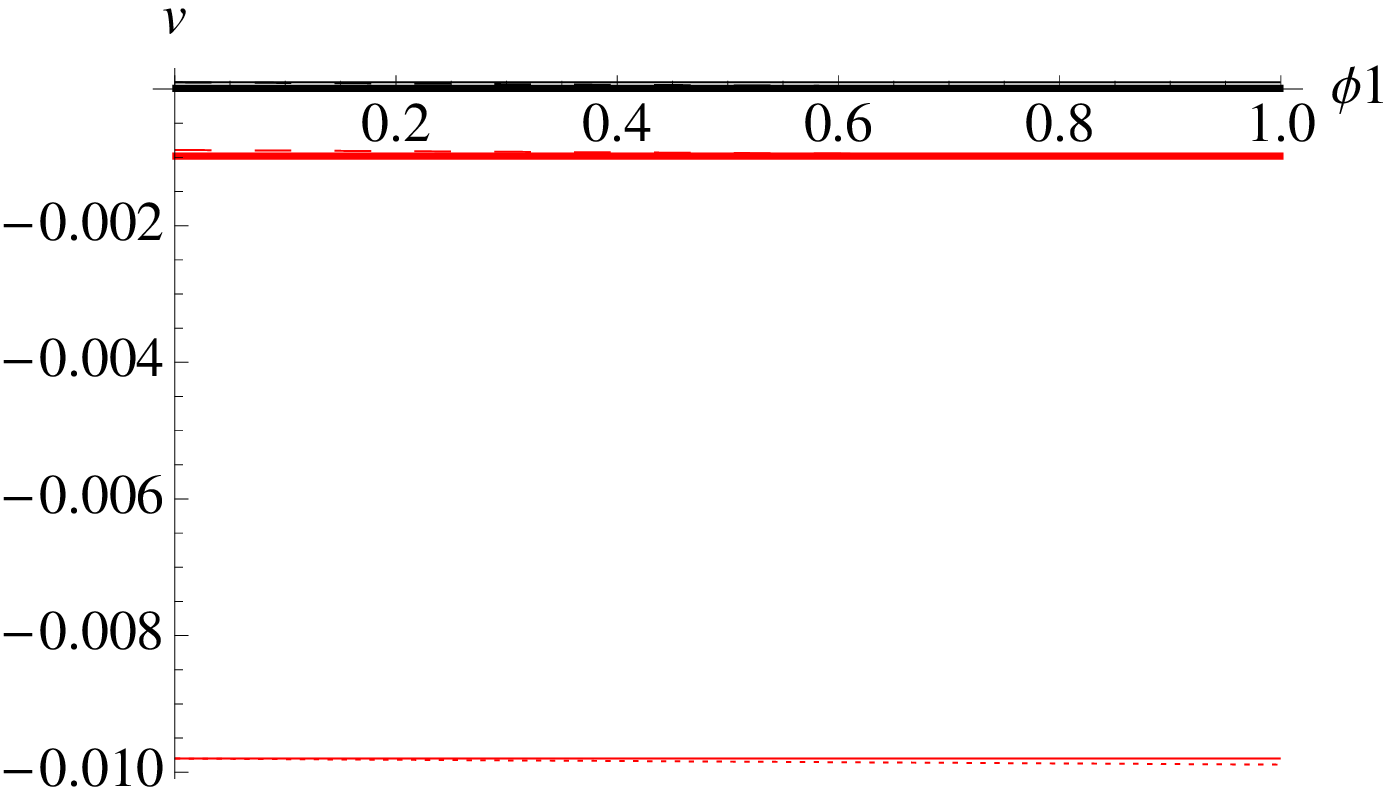}
   \hspace{0.5cm}
   \includegraphics[width=\columnwidth]{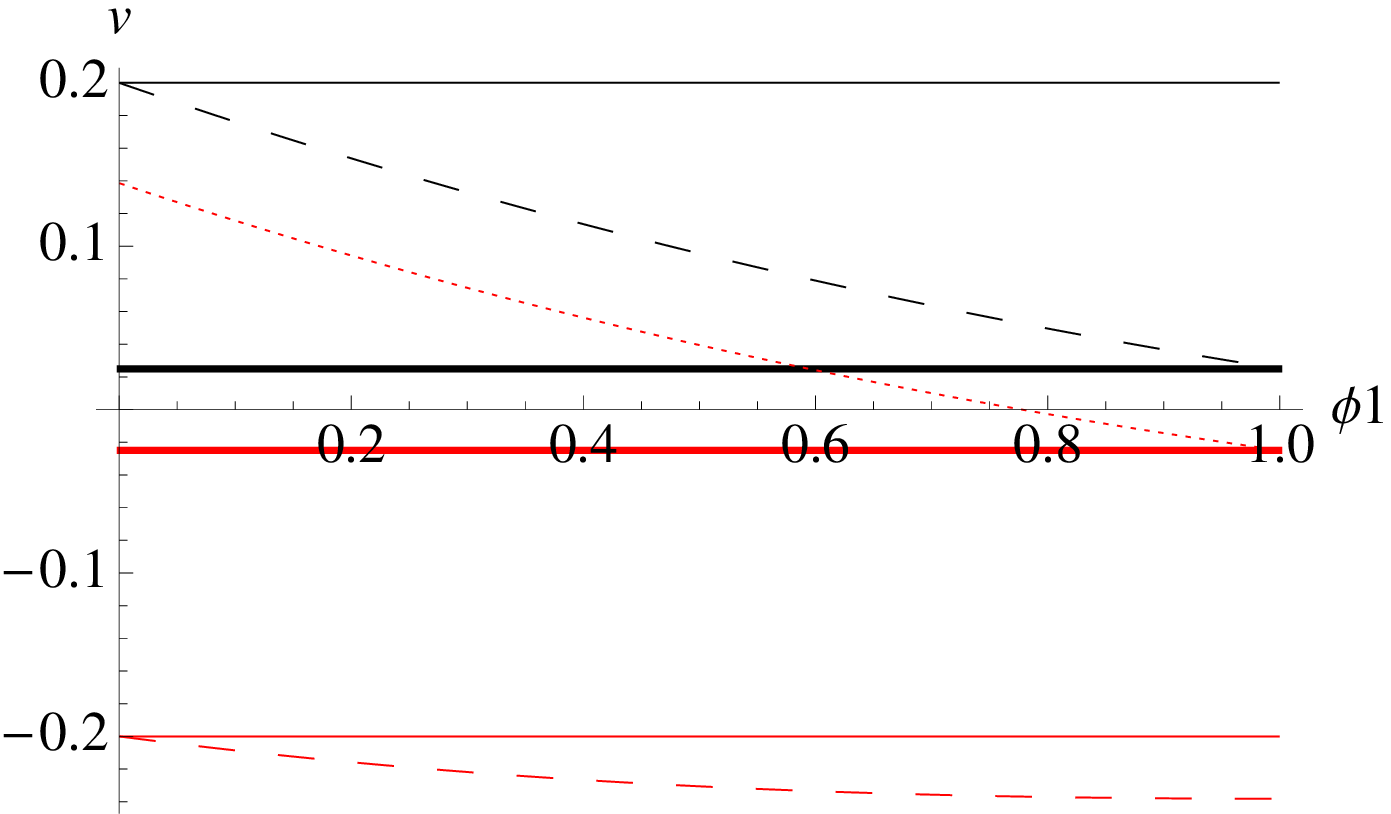} 
   \caption{Renormalised migration velocities as a function of the relative dust fraction $\phi_{1}$ in a two-dust-phase mixture. Positive velocities indicate outward migration. The left panel represents a typical initial situation for a protoplanetary disc. The parameters are set to $\rho_{0} = 1$, $\epsilon_{0} = 0.01$, $\Omega = 1$ and individual drag timescales $t_{1} = 10^{-3}$, $t_{2} = 10^{-2}$. The right panel mimics a situation where grains have grown and have concentrated due to settling ($\epsilon_{0} = 0.01$, $t_{1} = 0.1$, $t_{2} = 1$). The gas, the first (smaller grains) and the second (larger grains) dust phases are represented by black dashed, red dotted and red dashed lines respectively. As an indication, thin (thick) solid lines represent the gas and the dust velocities in a mixture made of the second (first) dust species only, i.e. $\phi_{1} = 0$ ($\phi_{1} = 1$). In the first configuration (left panel), grains are migrating inwards, in accordance with the single dust population case. Only a negligible dependence on the relative dust fraction is observed. In the second configuration (right panel), the smaller grains show outward migration when the relative dust fraction is $\lesssim 0.6$, with the velocity being of order the optimal migration velocity for grains in discs.}
   \label{fig:NSH86}
\end{center}
\end{figure*}
Valuable physical insight into the evolution of the system can be obtained by using the two dust phase population model described in Sect.~\ref{sec:twodust}.
When stationary equilibrium is reached, the radial velocities for the gas and the two dust species are:
\begin{equation}
\begin{pmatrix}
v_{\mathrm{g}x} \\
v_{\mathrm{d}1x}\\
v_{\mathrm{d}2x}
\end{pmatrix} =
\frac{\partial_{r}P}{\rhoz \left(1 - \epsilon \right)}
\begin{pmatrix}
t_{\mathrm{g}x} \\
t_{\mathrm{d}1x}\\
t_{\mathrm{d}2x}
\end{pmatrix} 
\end{equation}
where
\begin{eqnarray}
t_{\mathrm{g}x} & = &  \frac{\epsilon \left(1 - \epsilon \right) \left[ \phi_{1}t_{1} + \left(1 - \phi_{1} \right)t_{2} + \Omega^{2}\phi_{1}t_{1}t_{2}^{2} + \Omega^{2}\left( 1 -\phi_{1}\right)t_{2}t_{1}^{2}      \right]   }{ D\left(\Omega, \epsilon, \phi_{1},t_{1},t_{2} \right) } ,\label{eq:mig2gas}\\
t_{\mathrm{d}1x} & = & - \frac{\left(1 - \epsilon \right) \left[  \left(1 + \Omega^{2}t_{2}^{2}\right) t_{1} -\epsilon \left(\phi_{1} + \Omega^{2}t_{2}^{2} \right)t_{1}   - \epsilon\left(1 - \phi_{1} \right)t_{2}   \right] }{ D\left(\Omega, \epsilon, \phi_{1},t_{1},t_{2} \right) } ,\label{eq:mig2dust1}\\
t_{\mathrm{d}2x} & = & - \frac{\left(1 - \epsilon \right) \left[ \left(1 + \Omega^{2}t_{1}^{2} \right)t_{2}   - \epsilon\left(\left(1 - \phi_{1} \right) + \Omega^{2}t_{1}^{2} \right)t_{2} - \epsilon \phi_{1}t_{1}     \right]     }{ D\left(\Omega, \epsilon, \phi_{1},t_{1},t_{2} \right) } \label{eq:mig2dust2}.
\end{eqnarray}
and
\begin{eqnarray}
D\left(\Omega, \epsilon, \phi_{1},t_{1},t_{2} \right) & = & 1 +  \Omega^{2}\left[ \left(1 - \epsilon \phi_{1} \right)^{2}t_{1}^{2} + 2\epsilon^{2}\phi_{1}\left(1 - \phi_{1} \right)t_{1}t_{2} \right. \nonumber \\   
&& \left. + \left( 1 - \epsilon\left(1 - \phi_{1} \right) \right)^{2}t_{2}^{2} \right] + \left(1 - \epsilon \right)^{2}\Omega^{4}t_{1}^{2}t_{2}^{2} 
\label{eq:defD}
 \end{eqnarray}
As expected, Eqs.~\ref{eq:mig2gas}--\ref{eq:defD} are symmetric with respect to the transformation $\left(t_{1}\to t_{2} \right), \left(\phi_{1} \to 1 -\phi_{1} \right)$.  When the two dust populations degenerate ($t_{1}  =  t_{2}$, identical dust grains), Eqs.~\ref{eq:mig2gas}--\ref{eq:defD} reduce to
\begin{eqnarray}
t_{\mathrm{g}x} & = & \frac{\epsilon \left(1 - \epsilon \right)t_{1}}{1 + \Omega^{2}\left(1 - \epsilon \right)^{2}t_{1}^{2}}, \label{eq:mig1gas} \\
t_{\mathrm{d}x} & = &- \frac{ \left(1 - \epsilon \right)^{2} t_{1}}{1 + \Omega^{2}\left(1 - \epsilon \right)^{2}t_{1}^{2}} ,\label{eq:mig1dust}
\end{eqnarray}
which are the expressions obtained in the original derivation of \citet{NSH86} in the case $n =1$ (usually, the factor $\left( 1 - \epsilon\right) t_{1}$ is replaced by the stopping time $t_{\rm s}$). Dust loses angular momentum to the gas, implying that dust grains migrate inwards and the gas migrates outwards. Enforcing $\epsilon = 0$ directly in Eqs.~\ref{eq:mig2gas}--\ref{eq:defD} provides the usual expression of migration for individual isolated grains.

\subsubsection{Outward migration of dust particles}

Behaviours specific to multiple dust distributions are observed when the relative composition between the dust species is varied. In particular, an interesting limit consists of a situation where one of the two phases is infinitely diluted. As an example, we shall focus hereafter on the case $\phi_{1} \to 0$, since the two dust populations are symmetric. In this case, the inertia of the first dust phase is rigorously zero and grains behave like isolated individual particles. Thus, Eq.~\ref{eq:mig2dust1} reduces to
\begin{equation}
v_{\mathrm{d}1x} = - \frac{\left(1 - \epsilon \right) \left[ t_{1} -\epsilon t_{2} + \left(1 - \epsilon \right)\Omega^{2}t_{1}t_{2}^{2} \right]}{\left(1 + \Omega^{2}t_{1}^2 \right)\left(1 + \left(1 - \epsilon \right)^{2}\Omega^{2}t_{2}^{2} \right)} .
\label{eq:vdx1lim}
\end{equation}
As shown by Eq.~\ref{eq:vdx1lim}, $v_{\mathrm{d}1x}$ depends on $t_{1}$, but also on $t_{2}$ since the gas is dragged by the second dust species. The sign of $v_{\mathrm{d}1x}$ is given by the sign of the function $f_{\epsilon}$ defined by
\begin{equation}
f_{\epsilon} \left(T_{1}, T_{2} \right) = T_{1} - \epsilon T_{2} + \left(1 - \epsilon \right)T_{1}T_{2}^{2} ,
\label{eq:signvmig}
\end{equation}
where $T_{1}$ and $T_{2}$ are the individual Stokes numbers of each species, defined by $t_{1}\Omega = T_{1}$ and $t_{2}\Omega = T_{2}$. 

Fig.~\ref{fig:feps} summarises the detailed study of the function $f_{\epsilon}$. The important result is that for a range of values of $\left(T_{1}, T_{2} \right)$ which depends on the dust fraction $\epsilon$, $f_{\epsilon} < 0$, implying that the grains are migrating \textit{outwards}. Since this result is obtain at the limit $\phi_{1} \to 0$, it implies that there is a continuous range for increasing values of $\phi_{1}$ for which this result holds. \citet{Bai2010} observed this outwards migration for small grains as a result of their multiple grain size simulations. From Eq.~\ref{eq:signvmig}, outward migration occurs when
\begin{equation}
T_{1} < g_{\epsilon}\left(T_{1},T_{2} \right) = \frac{\epsilon T_{2}}{1 + \left(1 - \epsilon \right)T_{2}^{2}}.
\label{eq:condout}
\end{equation}
A necessary condition for this condition to be satisfied is (see Fig.~\ref{fig:feps}):
\begin{equation}
 T_{1} <g_{\epsilon}\left(T_{1},T_{2} \right) <  \epsilon T_{2} < T_{2}.
 \end{equation}
As a consequence, outward migration occurs only in the dust population with the smallest grain size, i.e. the one with the smallest value of $t_{k}$, which is the most efficiently dragged. Physically, the gas migrates outwards as an effect of the backreaction from the inward migration of the dense phase of large grains. Then, small dust grains efficiently couple to the gas and migrate outwards, rather than migrating inward as if it would be expected if they were the only dust population in the mixture. 

The largest possible value $T_{1\mathrm{c}}$ of outwardly migrating grains corresponds to the maximum of the function $g_{\epsilon}$. This is an increasing function of the dust fraction:
\begin{equation}
T_{1\mathrm{c}} = \frac{\epsilon}{2 \sqrt{1 - \epsilon}} ,
\label{eq:t1c}
\end{equation}
which is reached at $T_{2\mathrm{c}} = \left(1 - \epsilon \right)^{-1/2}$. Thus, at small values of $\epsilon$ (i.e. $\epsilon\ll1$), only very small grains can migrate outwards. However, when the dust-to-gas ratio becomes of order unity ($\epsilon \simeq 0.5$), $T_{1\mathrm{c}}$ becomes of order unity. Therefore, grains of intermediate size (i.e. having a Stokes number of order unity) can migrate outwards. In theory, very large values of $T_{1\mathrm{c}}$ can be reached in the limit $\epsilon \to 1$, but those regimes are not relevant for planet formation.

Fig.~\ref{fig:NSH86} compares the renormalised gas and dust velocities obtained for a single and two-dust-phase mixture as a function of the relative dust fraction $\phi_{1}$. Radial velocities are positive when the migration is outward. The parameters of the mixture are $\rho_{0} = 1$, $\epsilon_{0} = 0.01$, $\Omega = 1$ and $t_{1} = 10^{-3}$, $t_{2} = 10^{-2}$ (left panel) or  $\epsilon_{0} = 0.5$, $t_{1} = 0.1$, $t_{2} = 1$ (right panel). Those two sets of parameters are chosen to mimic a typical dust distribution in a protoplanetary disc before and after the growth and settling stage. In the first case, when the dust fraction is still small enough and the grains are small, each grains phase behaves almost as in the single grain case: particles migrate inwards with velocities that are almost identical to the ones found in the case $n = 1$. Corrections due to the change of relative dust composition are essentially negligible. More interesting is the case which mimics a stage where dust grains have grown and are highly concentrated in the disc mid plane. The presence of grains with Stokes number of order unity and dust-to-gas ratios of order unity (ie.. $\epsilon \simeq 0.5$) is expected (e.g. \citealt{BF2005,Zsom2011}). In such a situation, the right panel of Fig.~\ref{fig:NSH86} shows that outward dust migration occurs in this system. When larger grains ($t_{1} = 1$) dominate over the dust density ($\phi_{1} \lesssim 0.6 $), the smaller grains ($t_{2} = 0.1$) migrate outwards. A similar behaviour also occurs for any smaller grains in the first dust species ($t_{1}<0.1$). Importantly, the renormalised velocity of the outwardly migrating grains is $\simeq 0.1$, which is of order of the highest velocity which can be reached for inward migration in the case $n=1$, implying that this outwards drift can be quite efficient.

\subsubsection{Consequences for planet formation}

As discussed above, the maximum size of dust particles that can migrate outwards is an increasing function of the dust fraction. This result is of particular importance for planet formation. Initially, when dust grains are distributed over the entire disc, the dust fraction is of order  $\epsilon \simeq 10^{-3} - 10^{-2}$ and dust grains are micron-sized ($T_{2} \ll 1$). Thus, outward migration does not happen since it would only concern a population of non-physical (too small) grains ($T_{1} \lll 1$), whose migration efficiency would be negligible anyway. Then, grains grow and settle in the disc midplane where they concentrate. If a dust-to-gas ratio of order unity is reached, the presence of millimetre-sized grains ($T_{2} \simeq 1$, e.g. \citealt{Laibe2012}) can trigger outward migration of  hundredth-of-micron-sized grains (for which $T_{1} \simeq 0.1$) in a classical T Tauri star disc at 100AU. Such a scenario makes sense for real discs since the combination of growth and settling is known to provide particles with such sizes in the disc midplane \citep{Laibe2008, Brauer2008,Laibe2014c}. Moreover, the Stokes number is a decreasing function of the disc radius since it scales like the disc surface density (e.g. \citealt{Laibe2014b}). Therefore, on a global radial scale, outward migration of particles for which $S_{\rm t} \simeq 1$ bring grains to the outer regions of the disc, where their new Stokes number is larger than unity. This process therefore helps the particles to decouple from the gas and grow at this new location.

Outward migration of dust particles may explain the presence of large grains observed in the outer regions of protoplanetary discs \citep{Ricci2012}. When grains are initially growing, most of the dust mass is concentrated in the largest particles \citep{Blumwrum2008,Windmark2012b,Garaud2013}. This implies $\phi_{1} < 0.5$ and would support the outward migration mechanism detailed above.   However, this scenario would depend on the grain growth efficiency, which determines whether the density of the dust distribution is mostly concentrated into the small or the large grains.

All of this serves to reinforce \citet{Bai2010}'s remark that multiple dust phases should not be studied by treating the dust phases as if they were independently coupled to the gas. The outward migration of large grains found above would not be captured by such a procedure.

\begin{figure}
   \includegraphics[width=\columnwidth]{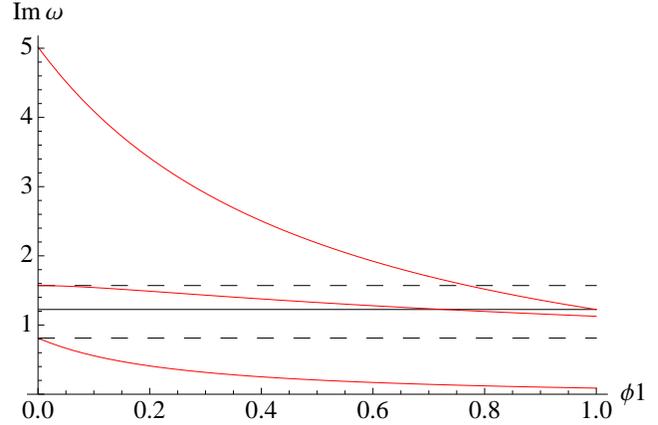}
   \caption{Growth rate of the unstable eigenmodes for the linear streaming instability problem in a two dust phases mixture as a function of the relative dust fraction $\phi_{1}$ (red solid lines). In this example, $\rho_{0} = 4$, $\epsilon_{0} = 0.75$, $t_{1} = 0.1$, $c_{\rm s} = 0.1$, $\eta = 0.05$, $_{x} = k_{z} = 30\pi /0.005$ and $t_{2} = 10$. The values of the limiting values in a single dust population limit are given by the dashed black ($\phi_{1} = 0$) and the solid black lines ($\phi_{1} = 1$). A small amount of strongly coupled dust grains can increase the efficiency of the instability by a factor three.}
   \label{fig:streaming}
\end{figure}

\subsubsection{A comment on the expression of the drag coefficients}

In a mixture with a single dust species ($n = 1$) it is convenient to denote the drag coefficient by the constant $K$. However, generalising this approach for multiple dust population using constant coefficients $K_{1}, K_{2}, ...,K_{n}$) instead of using the expression given by Eq.~\ref{eq:defKk}, would lead to incorrect results in the problem of the migration of multiple dust populations. Indeed, in the limit $\phi_{1}\to 0$ in the two-dust-species migration model studied above, the drag timescale $K_{1} / \left(\epsilon \phi_{1} \right)$ of the first dust species would tend to infinity instead of taking the finite value $t_{1}$, making the dust velocity incorrectly converge to the gas velocity. Hence, despite our earlier practice, we recommend use of the quantities $t_{k}$ rather than the drag coefficients.

\subsection{Linear growth of the streaming instability}

The streaming instability in dusty protoplanetary discs was discovered by \citet{Youdin2005}. It has since been studied in a number of papers (e.g. \citealt{Youdin2007,Johansen2007,Jacquet2011}) since it provides a mechanism to concentrate dust particles during the early stages of planet formation.  \citet{Youdin2005} showed that the stationary solution derived in Sect.~\ref{sec:NSH86} is unstable with respect to a linear perturbation that develops in the radial and the vertical direction simultaneously. The energy required for the amplification of the perturbation is provided by the background differential rotation between the phases. 

To perform a similar linear stability analysis in a multiple dust phase system, perturbations of the form $\delta \mathbf{A} = \delta \mathbf{A}_{0} e^{i\left(k_{x}x + k_{z}z - \omega t \right)}$ may be superimposed on the stationary solutions of the evolution equations for the mixture derived in Sect.~\ref{sec:NSH86}. The resulting linear system obtained is tediously long and of limited interest, and will not be reproduced here for clarity. However, two features of this system of equations are worth highlighting. Firstly, in a multiple fluid treatment of the gas and dust phases, perturbations of the convective derivatives of physical quantities give rise to terms of the form $\left[ -i\omega + i k_{x} v_{\mathrm{g}x0}\right] \delta \mathbf{A}_{0}$ (and similar expressions with $v_{\mathrm{dk}x0}$). In a one-fluid formalism, those are replaced by the simpler expression $-i \omega \delta \mathbf{A}_{0}$, since the stationary solution for the radial barycentric velocity of the mixture is identically zero. Secondly, in the one-fluid formalism, $\deltav_{\mathrm{g}0}$ and $\deltav_{\mathrm{d}k0}$ are first-order non-zero corrections to the background shear. This generates a large number of first order terms originating from the $\deltav$ contributions on the right-hand side of the evolution equations. Importantly, perturbations to the anisotropic pressure terms result in terms of order $\deltav_{0}^{2} \times \delta \mathbf{v}_{0}$ which are of \textit{third} order with respect to the background shear. This explains why the streaming instability is difficult to capture accurately in a global simulation of a protoplanetary disc, for which the background shear cannot be subtracted.

We have again employed the two phase dust mixture model of Sect.~\ref{sec:twodust} to gain a physical insight into the linear behaviour of the system, comparing its evolution to the limiting cases where either only the first or the second dust species are present in the mixture. Fig.~\ref{fig:streaming} shows the imaginary part of $\omega$ for the unstable modes of the linear system as a function of the relative dust fraction $\phi_{1}$. The following parameters are adopted: $\rho_{0} = 4$, $\epsilon_{0} = 0.75$, $t_{1} = 0.1$, $c_{\rm s} = 0.1$, $\eta = 0.05$ ($\eta$ being the dimensionless background radial pressure gradient), $k_{x} = k_{z} = 30\pi /0.005$, $t_{2} = 10$. If the dust phases degenerate into the single first dust phase ($\phi_{1} = 1$), the mixture reduces to the configuration of the linA mode described in \citet{Youdin2007}. Since $t_{2} > t_{1}$, the second dust phase adds grains that are individually less strongly coupled to the gas than those of the first phase.

As expected, the limit $\phi_{1} = 0$ and $\phi_{1} = 1$ generates the unstable mode obtained when only the second and the first dust species are present in the mixture, respectively.  Moreover, if $t_{1} = t_{2}$, the only unstable modes obtained are the ones of the corresponding single dust phase. In the general case, three unstable modes are found for the two-dust-species mixture, regardless of the value of $\phi_{1}$. The values of $\mathrm{Im}(w)$ of the unstable modes are monotonic functions of $\phi_{1}$. In this specific example, having $\phi_{1} \lesssim 0.75$ generates an unstable mode in the mixture which grows faster (i.e. up to a factor three in the limit $\phi_{1}\to 0$) than the modes generated by each dust species individually. Therefore, having a local dust distribution with multiple grain sizes can enhance the efficiency of planet formation in protoplanetary discs. Exploring the parameter space, we have not found a set of parameters that suppresses the streaming instability in a two dust phase mixture. Finally, in contrast to the \textsc{dustywave} problem, an analytic solution for the streaming instability can be found in the terminal velocity regime only for the case $n=1$ \citep{Youdin2005} since perturbations in $\delta \epsk$ do not add up to form a perturbation in $\delta \epsilon$ in the general case.

%=======================================================================================================
\section{Conclusion}
\label{sec:conclu}

We have derived a generalised formalism describing systems made of gas and any number of dust species as a single-fluid mixture, extending the approach developed in \citet{LP14a} towards realistic simulations of dusty astrophysical systems. This formalism brings three key advantages compared to a multiple-fluid approach:
\begin{enumerate}
\item It avoids the need for prohibitive spatial and temporal resolutions in order to correctly treat strongly coupled grains
\item It prevents the formation of artificial clumps which arise when dust particles concentrate below the gas resolution
\item It removes the need to interpolate between different phases in the numerical solution
\end{enumerate}

 We have derived the equations for the mixture in both Lagrangian and conservative Eulerian forms, for an arbitrary number of dust species as well as for continuous dust distributions, and in the zeroth and first-order approximations where the dust fractions are either constant or evolve according to diffusion equations, respectively. The main difference with multiple dust phases compared to the single dust phase mixture studied in \citet{LP14a} is that the differential velocities are related via a drag matrix. We have outlined in Sec.~\ref{sec:implicit} how these drag terms can be handled numerically using an implicit integration.

This single fluid formalism was then applied to both simple problems (the \textsc{dustybox}, the \textsc{dustyshock} and the \textsc{dustywave} problems) and more complex problems related to planet formation (grains radial-drift and streaming instability in protoplanetary discs). Where possible, analytic solutions for an arbitrary number of dust species have been derived. Where not, a two dust phase model was used to highlight the important physical mechanisms involved. 

As expected, the physics with multiple dust species is richer than with a single dust phase only. Several drag timescales are involved and the evolutions of physical quantities are not always monotonic. The global evolution of the mixture results from a balance between the relative strength of the drag terms and the relative mass in each dust phase.

The most interesting result concerns dust grains in protoplanetary discs. We find that after the growth and settling stage which concentrate the dust particles, large grains that are located in the outer disc regions can migrate outward. This would provide a simple explanation for the observed presence of (sub)millimetre-in-size grains at several tens if not hundreds of AU from their central star.

We also found that the presence of multiple grain sizes can increase the efficiency of the linear growth of the streaming instability. This would enhance planet formation in protoplanetary discs.

An obvious extension to the present work will be to translate this theoretical formalism into its SPH equivalent in order to solve the full non-linear system in three dimensions, generalising the study recently performed in \citet{LP14b}. Since the structure of the equations are similar to the single dust phase case, we expect this to be straightforward.

 Finally, the main limitation of the single fluid formalism at present is that is does not handle grain-grain interactions (in particular, grain growth and fragmentation). Addressing this issue is of tremendous importance but beyond the scope of the present paper.

\section*{Acknowledgments}
G. Laibe acknowledges funding from the European Research Council for the FP7 ERC advanced grant project ECOGAL. DJP is very grateful for funding via an ARC Future Fellowship, FT130100034, and Discovery Project grant DP130102078. G. Laibe thanks Ph. Bulois for a stimulating discussion and C. Clarke for the invitation to the IoA.

\begin{appendix}

\section{Properties of the spectrum of $\Odn$}
\label{app:spectrum}

To obtain a lower bound for the spectrum of $\Odn$ (or equivalently $\Wdn$), we first note that $\Wdn^{-1}$ is real and symmetric since $\Wdn$ is real and symmetric. $\Wdn^{-1}$ thus has positive real eigenvalues which are the inverse of $\Wdn$'s eigenvalues. Thus,
\begin{equation}
\lambda_{\min} ^{-1} < \mathrm{Tr} \left(\Wdn^{-1} \right) ,
\end{equation}
providing our lower bound for $\Odn$'s spectrum
\begin{equation}
\lambda_{\rm min} > \left( \sum_{k} \epsk \left(1 - \epsk \right)\tbk \right)^{-1} .
\end{equation}
A similar reasoning for the upper bound of $\Odn$'s spectrum would provide the following inequality:
\begin{equation}
\lambda_{\rm max} < \mathrm{Tr} \left(\Odn \right) = \sum_{k}\frac{1}{\dst t_{\mathrm{b}k}} \left( \frac{1}{\dst \epsilon_{k}  } +  \frac{1}{\dst \left(1 - \epsilon \right)  }\right) .
\label{eq:crap_upper}
\end{equation}
However, a better upper bound can be found by splitting the matrix $\Wdn$ according to
\begin{equation}
\Wdn = \mathrm{D} + \mathrm{U'} ,
\label{eq:decompo}
\end{equation}
where $D$ is the diagonal matrix defined in Eq.~\ref{eq:def_diag} and $\mathrm{U'}$ is the rank one matrix
\begin{equation}
\mathrm{U'}_{ij} = \frac{\sqrt{K_{i}K_{j}}}{\rho\left(1 - \epsilon \right)} = u'_{i} u^{' \mathrm{T}}_{i} ,
\end{equation}
where $u'$ is the vector defined by
\begin{equation}
u'_{i} = \sqrt{K_{i}}.
\end{equation}
$\mathrm{U}$ is a symmetric matrix whose unique eigenvalue $\lambda_{\mathrm{U'}}$ is given by its trace, i.e.
\begin{equation}
\lambda_{\mathrm{U}} = \dst \frac{\sum_{k} \tbk^{-1}}{\left(1 - \epsilon \right)} .
\end{equation}
Taking now advantage from the fact that the application which associates a symmetric matrix to its maximum eigenvalue is a norm, we apply the triangular inequality in Eq.~\ref{eq:decompo} and obtain
\begin{equation}
\lambda_{\rm max} \le \max_{k} \frac{1}{\epsk \tbk} + \dst \frac{\sum_{k} \tbk^{-1}}{\left(1 - \epsilon \right)} .
\label{eq:good_upper}
\end{equation}
Eq.~\ref{eq:good_upper} improves the upper bound given in Eq.~\ref{eq:crap_upper} by a factor $\mathcal{O}(1/n)$ when the $\epsk$ are small, which is likely to be the case in practice. Therefore:
\begin{equation}
\left( \sum_{k} \epsk \left(1 - \epsk \right)\tbk \right)^{-1} < \lambda_{\rm min} \le \lambda_{k} \le \lambda_{\rm max} \le \max_{k}\left(\frac{1}{\epsk\tbk} \right) + \frac{1}{\left(1 - \epsilon \right)}\sum_{k} \tbk^{-1} ,
\label{eq:bound_spectrum1}
\end{equation}
and using Eq.~\ref{eq:defKi2},
\begin{equation}
\left( \sum_{k} \left(1 - \epsk \right)\tk \right)^{-1} < \lambda_{\rm min} \le \lambda_{k} \le \lambda_{\rm max} \le \max_{k}\left(\frac{1}{\tk} \right) + \frac{1}{\left(1 - \epsilon \right)}\sum_{k} \epsk \tk^{-1} ,
\label{eq:bound_spectrum2}
\end{equation}

\end{appendix}

\bibliography{dustSPH}

\label{lastpage}
\end{document}

%% file: journaux.tex
%
%  These Macros are taken from the AAS TeX macro package version 4.0.
%  Include this file in your LaTeX source only if you are not using
%  the AAS TeX macro package and need to resolve the macro definitions
%  in the BibTeX entries returned by the ADS abstract service.
%
%  For more information on the AASTeX macro package, please see the URL
%	http://www.aas.org/publications/aastex.html
%  For more information about ADS abstract server, please see the URL
%	http://adswww.harvard.edu/ads_abstracts.html
%

% Abbreviations for journals.  The object here is to provide authors
% with convenient shorthands for the most "popular" (often-cited)
% journals; the author can use these markup tags without being concerned
% about the exact form of the journal abbreviation, or its formatting.
% It is up to the keeper of the macros to make sure the macros expand
% to the proper text.  If macro package writers agree to all use the
% same TeX command name, authors only have to remember one thing, and
% the style file will take care of editorial preferences.  This also
% applies when a single journal decides to revamp its abbreviating
% scheme, as happened with the ApJ (Abt 1991).

\def\jnl@style{\it}
%commente par Seb
\def\aaref@jnl#1{{\jnl@style#1}}
%ref remplace par aaref pour eviter conflit...

\def\aaref@jnl#1{{\jnl@style#1}}

\def\aj{\aaref@jnl{AJ}}                   % Astronomical Journal
\def\araa{\aaref@jnl{ARA\&A}}             % Annual Review of Astron and Astrophys
\def\apj{\aaref@jnl{ApJ}}                 % Astrophysical Journal
\def\apjl{\aaref@jnl{ApJ}}                % Astrophysical Journal, Letters
\def\apjs{\aaref@jnl{ApJS}}               % Astrophysical Journal, Supplement
\def\ao{\aaref@jnl{Appl.~Opt.}}           % Applied Optics
\def\apss{\aaref@jnl{Ap\&SS}}             % Astrophysics and Space Science
\def\aap{\aaref@jnl{A\&A}}                % Astronomy and Astrophysics
\def\aapr{\aaref@jnl{A\&A~Rev.}}          % Astronomy and Astrophysics Reviews
\def\aaps{\aaref@jnl{A\&AS}}              % Astronomy and Astrophysics, Supplement
\def\azh{\aaref@jnl{AZh}}                 % Astronomicheskii Zhurnal
\def\baas{\aaref@jnl{BAAS}}               % Bulletin of the AAS
\def\jrasc{\aaref@jnl{JRASC}}             % Journal of the RAS of Canada
\def\memras{\aaref@jnl{MmRAS}}            % Memoirs of the RAS
\def\mnras{\aaref@jnl{MNRAS}}             % Monthly Notices of the RAS
\def\pra{\aaref@jnl{Phys.~Rev.~A}}        % Physical Review A: General Physics
\def\prb{\aaref@jnl{Phys.~Rev.~B}}        % Physical Review B: Solid State
\def\prc{\aaref@jnl{Phys.~Rev.~C}}        % Physical Review C
\def\prd{\aaref@jnl{Phys.~Rev.~D}}        % Physical Review D
\def\pre{\aaref@jnl{Phys.~Rev.~E}}        % Physical Review E
\def\prl{\aaref@jnl{Phys.~Rev.~Lett.}}    % Physical Review Letters
\def\pasp{\aaref@jnl{PASP}}               % Publications of the ASP
\def\pasj{\aaref@jnl{PASJ}}               % Publications of the ASJ
\def\qjras{\aaref@jnl{QJRAS}}             % Quarterly Journal of the RAS
\def\skytel{\aaref@jnl{S\&T}}             % Sky and Telescope
\def\solphys{\aaref@jnl{Sol.~Phys.}}      % Solar Physics
\def\sovast{\aaref@jnl{Soviet~Ast.}}      % Soviet Astronomy
\def\ssr{\aaref@jnl{Space~Sci.~Rev.}}     % Space Science Reviews
\def\zap{\aaref@jnl{ZAp}}                 % Zeitschrift fuer Astrophysik
\def\nat{\aaref@jnl{Nature}}              % Nature
\def\iaucirc{\aaref@jnl{IAU~Circ.}}       % IAU Cirulars
\def\aplett{\aaref@jnl{Astrophys.~Lett.}} % Astrophysics Letters
\def\apspr{\aaref@jnl{Astrophys.~Space~Phys.~Res.}}
                % Astrophysics Space Physics Research
\def\bain{\aaref@jnl{Bull.~Astron.~Inst.~Netherlands}} 
                % Bulletin Astronomical Institute of the Netherlands
\def\fcp{\aaref@jnl{Fund.~Cosmic~Phys.}}  % Fundamental Cosmic Physics
\def\gca{\aaref@jnl{Geochim.~Cosmochim.~Acta}}   % Geochimica Cosmochimica Acta
\def\grl{\aaref@jnl{Geophys.~Res.~Lett.}} % Geophysics Research Letters
\def\jcp{\aaref@jnl{J.~Chem.~Phys.}}      % Journal of Chemical Physics
\def\jgr{\aaref@jnl{J.~Geophys.~Res.}}    % Journal of Geophysics Research
\def\jqsrt{\aaref@jnl{J.~Quant.~Spec.~Radiat.~Transf.}}
                % Journal of Quantitiative Spectroscopy and Radiative Transfer
\def\memsai{\aaref@jnl{Mem.~Soc.~Astron.~Italiana}}
                % Mem. Societa Astronomica Italiana
\def\nphysa{\aaref@jnl{Nucl.~Phys.~A}}   % Nuclear Physics A
\def\physrep{\aaref@jnl{Phys.~Rep.}}   % Physics Reports
\def\physscr{\aaref@jnl{Phys.~Scr}}   % Physica Scripta
\def\planss{\aaref@jnl{Planet.~Space~Sci.}}   % Planetary Space Science
\def\procspie{\aaref@jnl{Proc.~SPIE}}   % Proceedings of the SPIE

\let\astap=\aap
\let\apjlett=\apjl
\let\apjsupp=\apjs
\let\applopt=\ao